\documentclass{aa}  
\usepackage{graphicx}
\usepackage{natbib}
\usepackage{txfonts}
\usepackage{multirow}
\begin{document}
   \title{Multi-wavelength observations of Proxima Centauri \thanks{Based on observations
collected at the European Southern Observatory, Paranal, Chile, 082.D-0953A and on observations
obtained with \emph{XMM-Newton}, an ESA science mission with instruments and contributions directly funded 
by ESA Member states and NASA.}\fnmsep\thanks{Full Table \ref{linetable} is only available in electronic form at the CDS}}

   \author{B. Fuhrmeister$^{1}$, S. Lalitha$^{1}$, K. Poppenhaeger$^{1}$, N. Rudolf$^{1}$,
           C. Liefke$^{1,2}$,\\ A. Reiners$^{3}$, J. H. M. M. Schmitt$^{1}$, \and J.-U. Ness$^{4}$
          }
   \authorrunning{B. Fuhrmeister et al.}
   
   \offprints{B. Fuhrmeister}

   \institute{$^{1}$Hamburger Sternwarte, University of Hamburg,
              Gojenbergsweg 112, 21029 Hamburg, Germany\\
              $^{2}$Zentrum f\"ur Astronomie, M\"onchhofstra\ss e 12-14, 69120 Heidelberg, Germany\\
              $^{3}$Institute of Astrophysics, University of G\"ottingen, 
              Friedrich-Hund-Platz 1, 37077 G\"ottingen, Germany\\
              $^{4}$XMM-Newton Science Operations Centre, European Space Agency (ESA / ESAC), 28691 Villanueva de la Ca\~{n}ada, Madrid, Spain\\
              \email{bfuhrmeister@hs.uni-hamburg.de}
               }

   \date{Received XXXX; accepted XXXX}

 
\abstract
   {}
{We report simultaneous observations of the nearby flare star Proxima Centauri with VLT/UVES 
and \emph{XMM-Newton} over three nights in March 2009. 
Our optical and X-ray observations cover the star's quiescent state, as
   well as its flaring activity and allow us to probe the stellar atmospheric conditions  
   from the photosphere into the chromosphere, and then the corona during its different activity stages. }
   {Using the X-ray data, we investigate variations in coronal densities and abundances and infer loop
    properties for an intermediate-sized flare. The optical data are used to investigate the magnetic 
    field and its possible variability, to construct an emission line list for the chromosphere, and use certain 
  emission lines to construct physical models of Proxima Centauri's 
chromosphere.}
   {We report the discovery of a weak optical forbidden \ion{Fe}{xiii} line at 3388 \AA\, during the more 
    active states of Proxima Centauri. For the intermediate flare, we find two secondary flare events
    that may originate in neighbouring loops, and discuss the line asymmetries observed during 
this flare in \ion{H}{i}, \ion{He}{i}, and \ion{Ca}{ii} lines. 
The high time-resolution in the H$\alpha$
    line highlights strong temporal variations in the observed line asymmetries, 
    which re-appear during a secondary flare event. 
We also present theoretical modelling with the stellar atmosphere code PHOENIX to construct flaring chromospheric models.}
   {}


   \maketitle
%

\section{Introduction}

Coronal heating of cool stars is thought to be driven by magnetic fields generated
in the stellar convection zones. For stars from spectral type F to mid-M,
the magnetic field production process
can be modelled with an $\alpha\Omega$ dynamo \citep{Parker1955}. However, stars of
spectral type M3 or later  are fully convective \citep{Chabrier, Dorman} and are not expected to
undergo the same dynamo process as earlier-type stars \citep{Browning}. Other processes such
as turbulent or $\alpha^2$ dynamos should not be as effective as the
$\alpha\Omega$ dynamo and, indeed, many late M dwarfs have rather low X-ray
luminosities in their states of quiescence \citep{Robrade2005}. Nevertheless,
the same stars are
capable of producing strong flares of short as well as long
durations \citep{Stelzer2006, Robrade2010}, so there must be some mechanism to allow violent releases of large amounts of magnetic energy. 
Magnetic flux densities are enormous in mid- to 
late-M stars, providing firm evidence of an 
efficient mechanism to produce and maintain strong magnetic fields 
(e.g.\ \citet{Reiners2007,Reiners2010}). In summary, the magnetic activity
processes in late M dwarfs are not yet
thoroughly understood in detail. Multi-wavelength observations cover different
activity indicators and can therefore help us to reconstruct a picture of the whole
stellar atmosphere. Examples of multi-wavelength observations can be found in e.\,g.
\citet{Osten}, \citet{Berger}, \citet{Kowalski_2}, and \citet{CNLeoflare}.

The energy released in flares is thought to originate from the interaction of magnetic
fields in the convective zone of the star and photospheric motions that
entangle the magnetic loops penetrating the stellar surface. The
energy release is triggered by magnetic instabilities that cause the 
entangled magnetic field lines to reconnect in the corona leading to 
heating and particle acceleration in the reconnection region. Particles are 
accelerated downwards into the chromosphere where they collisionally heat the 
denser plasma, which in turn expands and evaporates into the corona. The 
coronal loop is filled by dense, hot plasma, which is then 
observable as a flare in soft X-rays (see e.\,g. \citet{Haisch1991}). 

Proxima Centauri, with a distance of only $1.3$~pc \citep{Leeuwen} the star closest 
to the Sun, is a magnetically active star of spectral type dM5.5.
It has been frequently observed by various X-ray satellites: {\em Einstein} \citep{Einstein}, 
{\em EXOSAT-IUE} \citep{IUE}, {\em XTE} \citep{RXTE}, {\em ROSAT} \citep{Voges} , {\em ASCA} \citep{ASCA}, 
{\em Chandra} \citep{Wargelin}, and {\em XMM-Newton} \citep{Guedel_ProxCen_2}. 
Its quiescent X-ray luminosity varies in the range  $L_X\approx
(4-16)\times 10^{26}$~erg\,s$^{-1}$, which is comparable to that of 
the Sun despite its
$50$ times smaller surface area. Over the past $30$ years,
several X-ray flares of Proxima Centauri have been observed, with
the most extreme peak luminosities observed in a 2001 {\em XMM-Newton}
observation \citep{Guedel_ProxCen_2}, which exceeded typical quiescent state
X-ray fluxes by a factor of $\approx 100$. 
Here we present new data for Proxima Centauri's coronal and chromospheric properties,
together with simultaneous measurements of its large-scale magnetic
field strength.

Our paper is structured as follows.
In Section~2, we describe our observations obtained with VLT/UVES and {\em XMM-Newton}. 
In Section~3, we compare the timing behaviour of Proxima Centauri in different wavelength bands. 
The coronal properties of Proxima Centauri such as temperatures and elemental abundances are 
presented in Section~4, while Section~5 describes the chromospheric and transition 
region properties of the star. Section~6 and 7 contain a discussion of the presented
 findings and our conclusions.


\section{Observations and data analysis}

The multi-wavelength observations reported in this paper were obtained strictly
simultaneously with \emph{XMM-Newton} and ESO's Kueyen telescope equipped
with the Ultraviolet-Visual Echelle Spectrograph (UVES) on 9, 11, and 13
March 2009 (labelled ''night~1'', ''night~2'' and ''night~3'' in the following; see also Table~\ref{obsids}).
Due to the proximity of Proxima Centauri, interstellar absorption is negligible for the optical as well as for the X-ray data.

\subsection{Optical UVES data}\label{UVES_observations}

For our optical observations, the UVES spectrograph was operated in a
dichroic mode leading to a spectral coverage from about 3290~\AA\, to 4500~\AA\, in 
the blue arm and 6400~\AA\, to 10080~\AA\, in the red arm with a small gap from 
8190~\AA\, to 8400~\AA\, caused by the CCD mosaic\footnote{A detailed description of the UVES 
spectrograph is available at http://www.eso.org/instruments/uves/doc/}. For the red arm, a non-standard setup 
was used to ensure coverage of the H$\alpha$ line.
We used exposure times varying from 1000 
to 1800 seconds for the blue arm and of 90 to 450 seconds in the red arm
due to variable seeing conditions.
In the blue arm, we obtained 24, 17, and 15 useful spectra 
during the three nights and in the red arm  215, 168, and 179 useful spectra, respectively.
The typical resolution of our spectra is $\sim$ 45\,000.
The red arm spectra were reduced using the UVES pipeline vers. 4.3.0 \citep{UVESpipeline}\footnote{The UVES pipeline manual can be found at ftp.eso.org/pub/dfs/pipelines/uves-cpl/uves-pipeline-manual-13.0.pdf}.
 The blue arm spectra could not be properly reduced with the pipeline software but 
had to be reduced manually using the IDL 
reduction software {\tt REDUCE} \citep{reduce} for several reasons. 
To obtain a simultaneous coverage with \emph{XMM-Newton},
the star had to be observed at rather high airmass starting with a maximum of 2.4.
Since the positioning of the star was accomplished with the red arm, a wider slit
 had to be used for the  blue spectra, which resulted in an overlap
 of the bluest spectral orders. Therefore no dark exposure
could be acquired and the flat-field and science spectra extraction had to be carried out 
with a fixed width and without scattered light. To complicate the spectral 
reduction even more, quite a number of spectra are contaminated by solar stray-light
(from the Moon),
which is outshone by the star in the red part of the spectrum but 
can be clearly recognised in the blue part of the spectrum.

The wavelength calibration in the blue arm was carried out with thorium-argon spectra
with an accuracy of $\sim$ 0.04 \AA. Since the weather conditions and especially
the seeing varied throughout the nights, no absolute flux calibration with a standard star could be performed. 
To obtain an {\it a posteriori} flux calibration, we used synthetic stellar spectra provided by the stellar
atmosphere program PHOENIX \citep{phoenix}. We determined the best-fit
stellar model using a grid with $T_{\mathrm{eff}}$ varying from 2700~K to 3400~K
in steps of 100~K and log~$g$ of 4.5, 5.0, and 5.5 and found that
model to have $T_{\mathrm{eff}}= 3100$ K and log~$g=5.5$ in good agreement with
the values derived by \citet{Demory}, who had derived $T_{\mathrm{eff}}= 3098~\pm$ 56~K and log~$g=5.2$ using the VLTI. 
For the blue spectra, only data at wavelength redward of 4000 \AA\, were used
for the calibration since at short wavelengths the data are dominated by chromospheric emission
and thus the model spectra are  unreliable. We estimate that our errors in the flux calibration in the blue band 
are about a factor of two. A comparison with a flux-calibrated
spectrum of Proxima Centauri acquired by \citet{Cincunegui} and \citet{Cincunegui_2} shows that our fluxes
are higher by about a factor of two for the blue band and by a factor of two to three in the red band.
Since these authors used a standard star and low resolution spectra
for their flux calibration, the main source of error in our calibration seems to be the model spectrum. 

In addition to the spectral data, we obtained a blue and a red light curve 
using the UVES exposure-meters, i.\,e., one blue and
red photometer located in the two arms of the spectrograph. These data are normally obtained for engineering
purposes only and are not flux calibrated. We note that the flux bands of the exposure-meters 
are not identical to the spectral bands, the blue flux band in particular being ''redder''
than the blue spectrum.

\subsection{X-ray data}

Each of our VLT observations was accompanied by three simultaneous 30~ks 
observations conducted with \emph{XMM-Newton}; the exact observation times 
with the ObsIDs being 
given in Table \ref{obsids}. On board \emph{XMM-Newton},
there are three telescopes focusing X-rays onto three CCD cameras (one PN and two metal oxide semi-conductor (MOS) cameras with a sensitivity
range of $\approx 0.2-15$~keV), which together form the European
Photon Imaging Camera (EPIC). The X-ray telescopes with the MOS detectors are also
equipped with reflection gratings. The two RGS (Reflection
Grating Spectrometers) provide high-resolution X-ray spectroscopy
(E/$\Delta$E $\approx$ 200-800) in the energy range of 0.35-2.5~keV capable of
resolving individual X-ray emission lines. The X-ray instruments are accompanied by the
Optical Monitor (OM), an optical/UV telescope that can be used with
different filters for imaging and time-resolved photometry. Useful data of Proxima 
Centauri were obtained with the OM, EPIC, and RGS detectors, which were all operated simultaneously. The PN and MOS detectors were operated with the
medium filter in full frame and large window mode, respectively.  The OM
was operated in fast mode with $0.5$~s cadence using
the U~band filter covering a band pass of $300-390$~nm.

All XMM-Newton data were reduced using the standard
XMM-Newton Science Analysis System (SAS) software, version 10.0. The EPIC
light curves and spectra were obtained using standard filtering criteria. 
We adopted an extraction radius of $15''$ centred on the source; 
for background subtraction we used nearby source-free areas. Source counts with energies in the energy range 0.2 -- 10~keV were considered for 
the scientific analysis. For the first two nights, the X-ray observations displayed only a low background level. 
During the third night, short phases with high background levels were present; these time intervals were excluded 
from our spectral analysis. The X-ray light curves are background-subtracted and binned by 100~s unless stated otherwise.

\begin{table}
\caption{\label{obsids} The three XMM-Newton observations of Proxima Centauri in March 2009.}
\begin{tabular}[htbp]{cccc}
\hline
\hline
ObsID & start time & duration (ks) & label \\
\hline \\
0551120301 &  2009-03-10 02:23:34 & 28.7 & night 1\\
0551120201 &  2009-03-12 02:11:19 & 30.7 & night 2\\
0551120401 &  2009-03-14 02:20:45 & 28.7 & night 3\\
\hline
\end{tabular}
\end{table}

Spectral analysis was carried out with Xspec V12.5.0 \citep{xspec} for
the overall fitting processes and CORA \citep{cora} for fitting of individual spectral lines. 
For the overall analysis in Xspec, we used models with several temperature components
assuming the same elemental abundance for each
component. These models are based on a collisionally ionised
optically thin gas  calculated with the APEC code
\citep{apec}. Abundances are calculated relative to the solar photospheric
values given by \citet{Grevesse}.




 \section{Multi-wavelength timing behaviour}

In Figures \ref{lightcurve1}, \ref{lightcurve2}, and \ref{lightcurve3}, we present the background-subtracted X-ray light
curves (taken with EPIC-PN) and the OM U-band light curve for our three Proxima Centauri 
observations, as well as the optical UVES blue-band flux and several chromospheric
line fluxes; the X-ray light curves are binned in units of 100s, and data for the OM in units of 10s, while 
the UVES exposure-meter data are binned in units of 5s.

\subsection{Quiescence and small flares}\label{quasi-quiescent}

\begin{figure}
\begin{center}
\includegraphics[width=8cm,trim=0mm 11mm 0mm 0mm,clip]{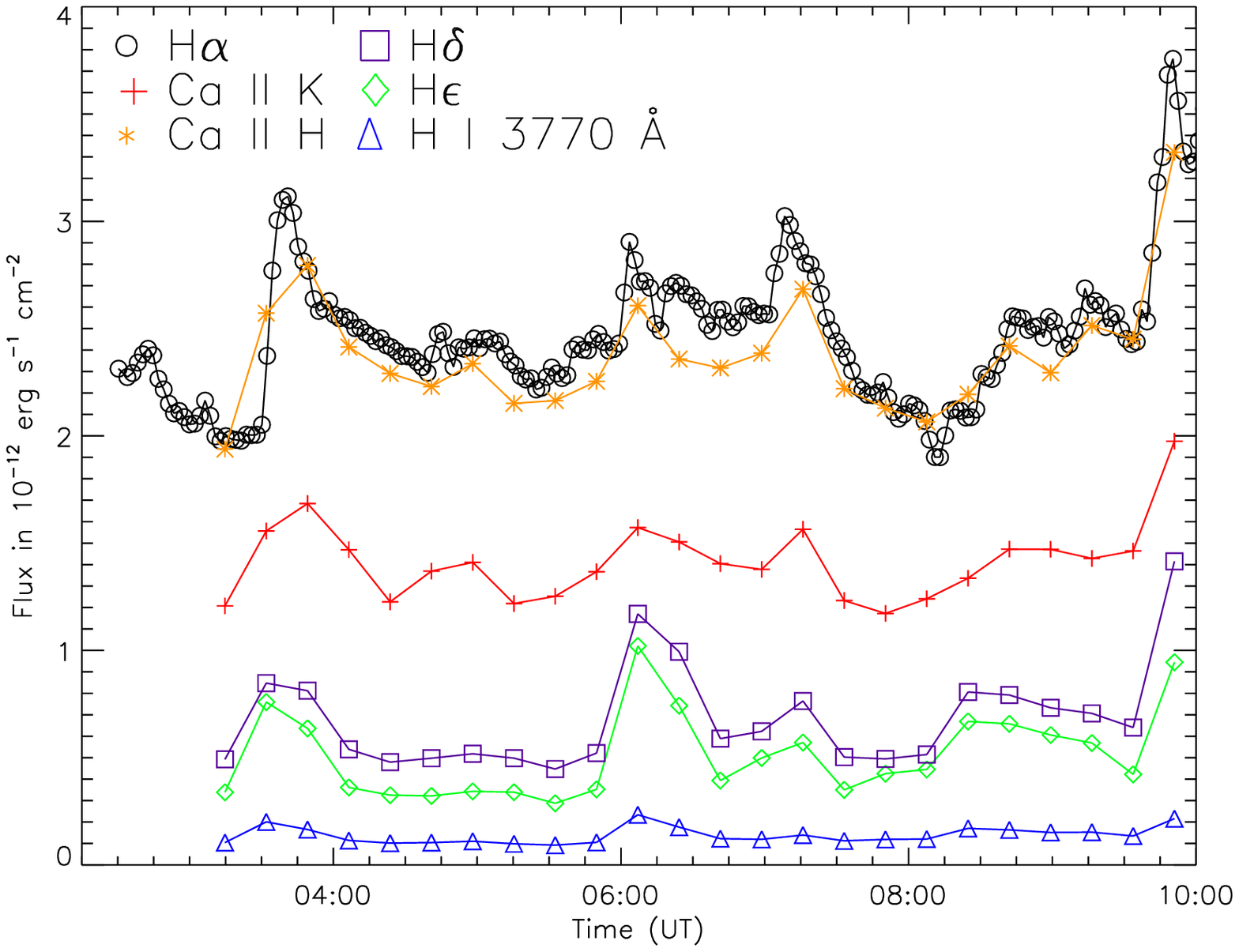}\vspace{-0.4mm}
\includegraphics[width=8cm]{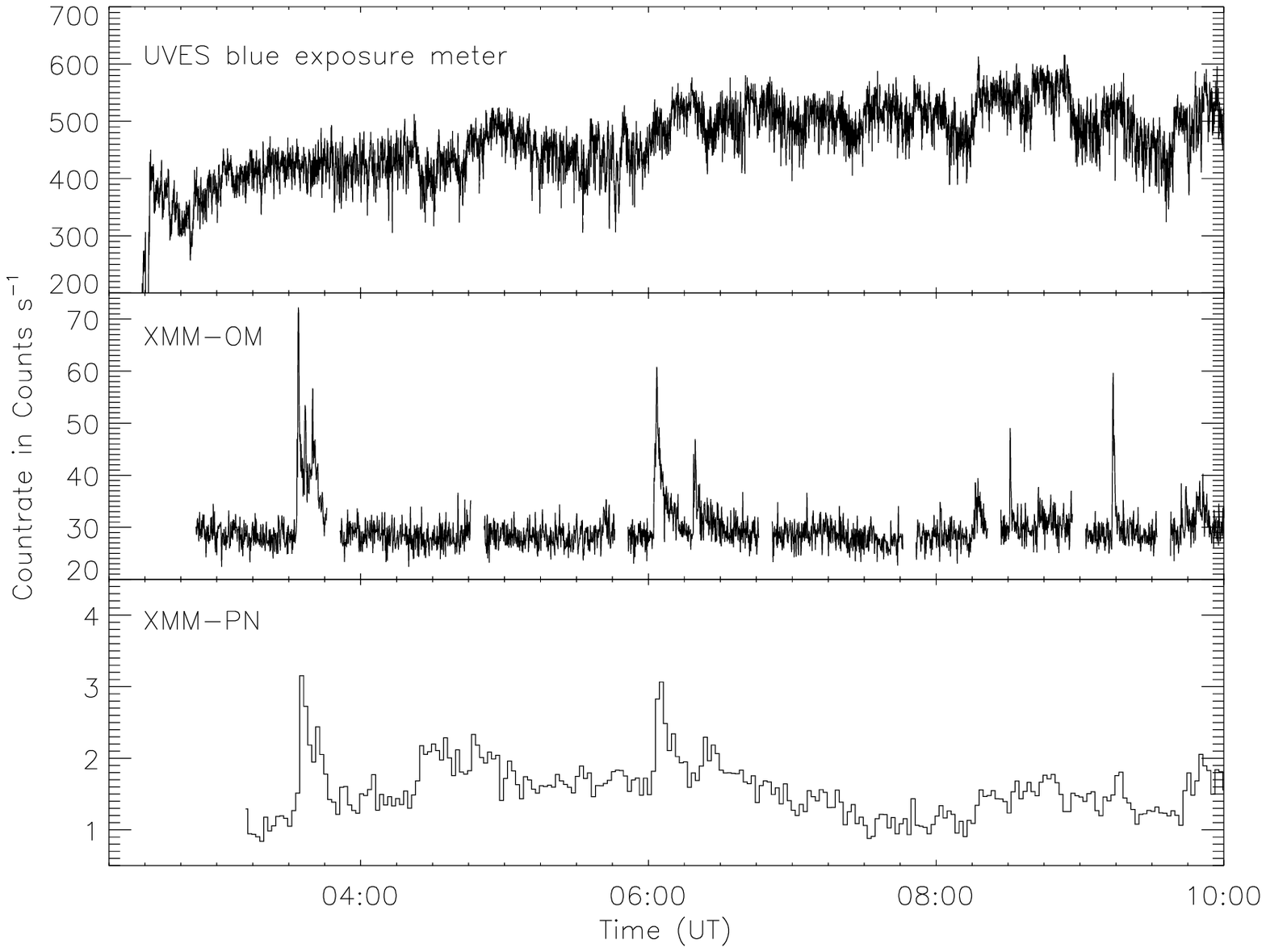}
\caption{\label{lightcurve1} Light curves of Proxima Centauri from night 1 as seen in the fluxes of 
characteristic chromospheric emission lines (top, note the better time binning provided by the H$\alpha$ line) 
as well as optical UVES, OM, and X-ray light curves (bottom).}
\end{center}
\end{figure}

\begin{figure}
\begin{center}
\includegraphics[width=8cm,trim=0mm 11mm 0mm 0mm,clip]{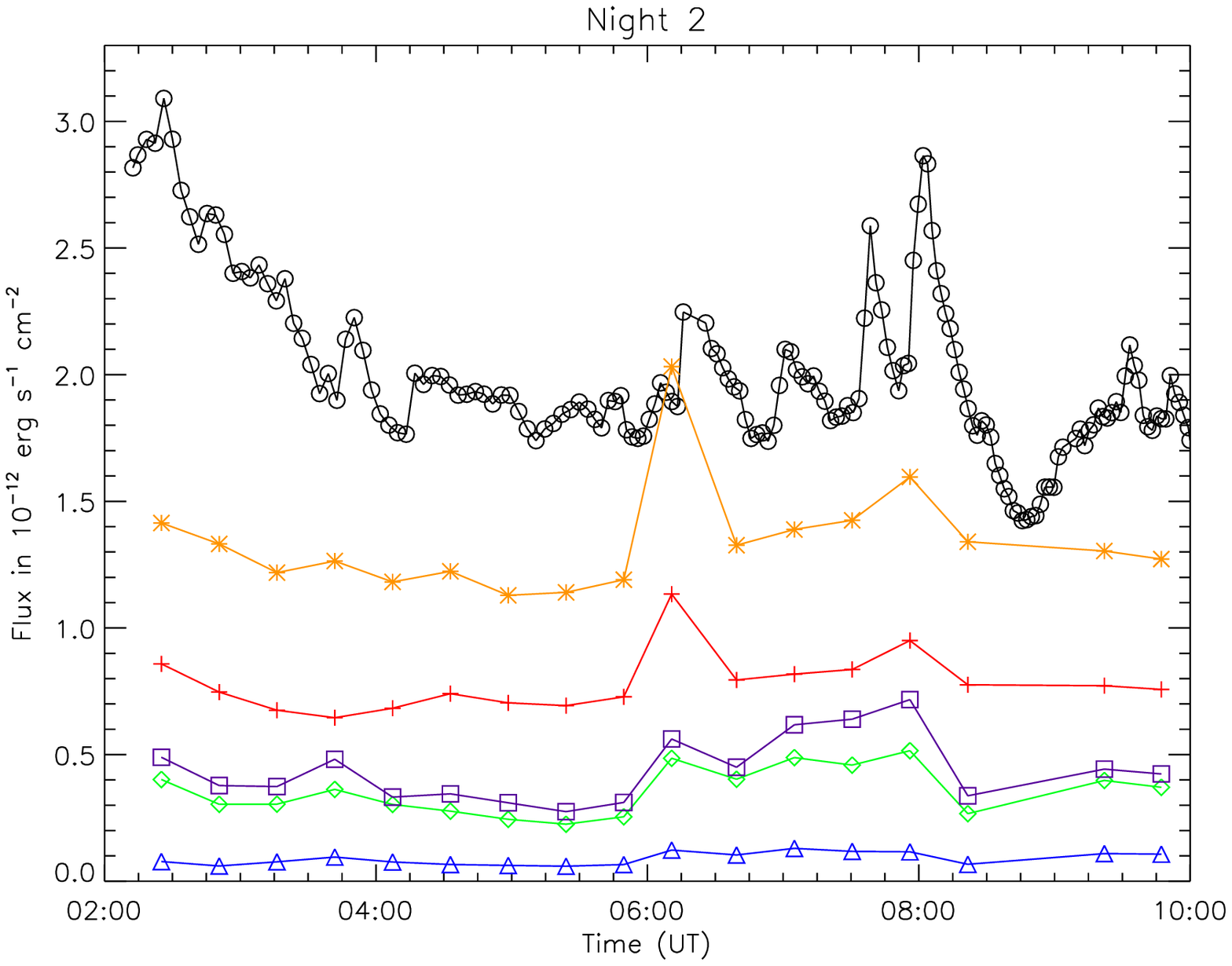}\vspace{-0.4mm}
\includegraphics[width=8cm]{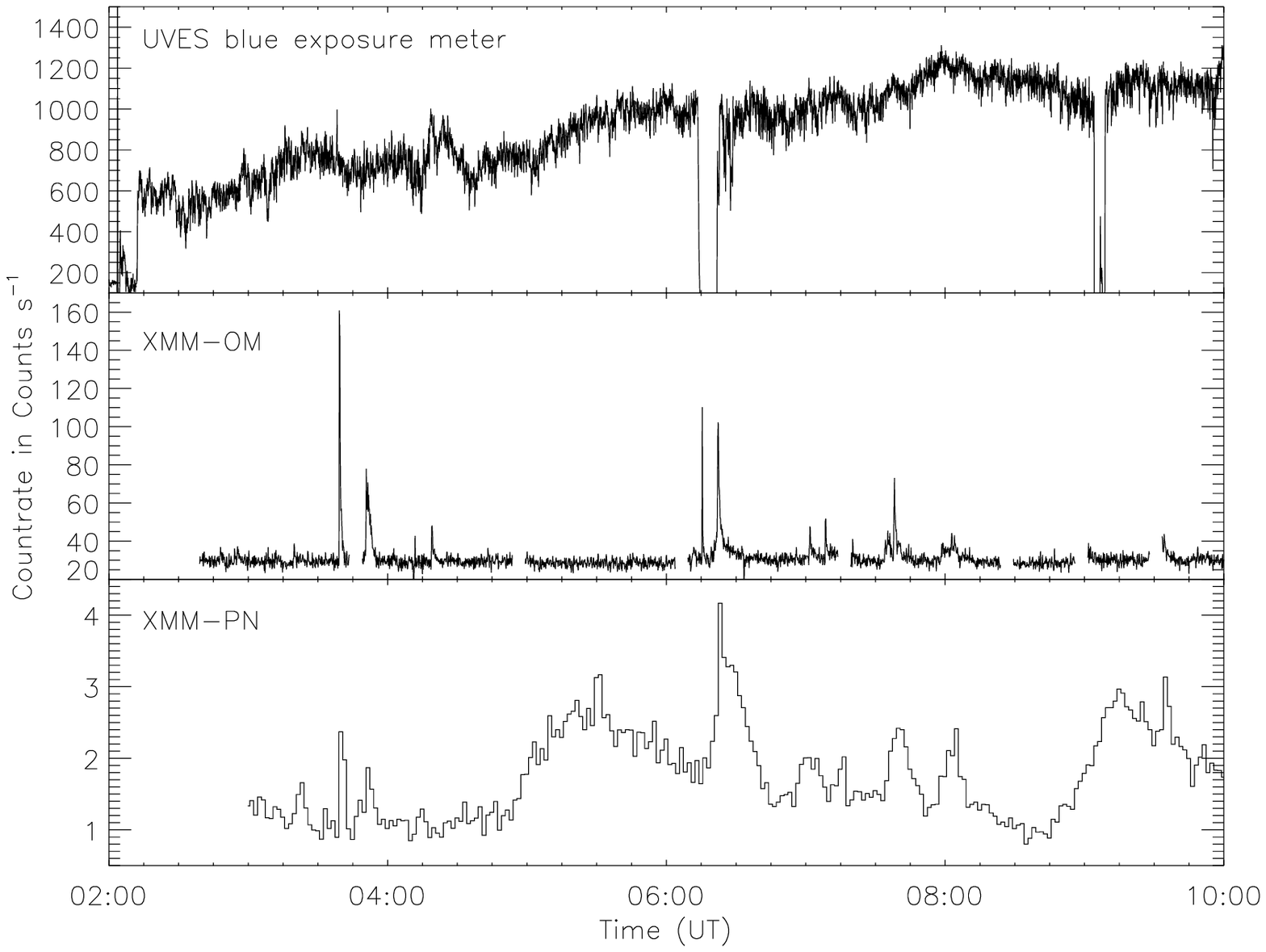}
\caption{\label{lightcurve2} As Fig.~\ref{lightcurve1}, but for night 2. The two gaps in the UVES 
light curve are due to technical problems that led to the loss of the guide star. The legend is the same as
in Fig. \ref{lightcurve1}.}
\end{center}
\end{figure}

\begin{figure}
\begin{center}
\includegraphics[width=8cm,trim=0mm 11mm 0mm 0mm,clip]{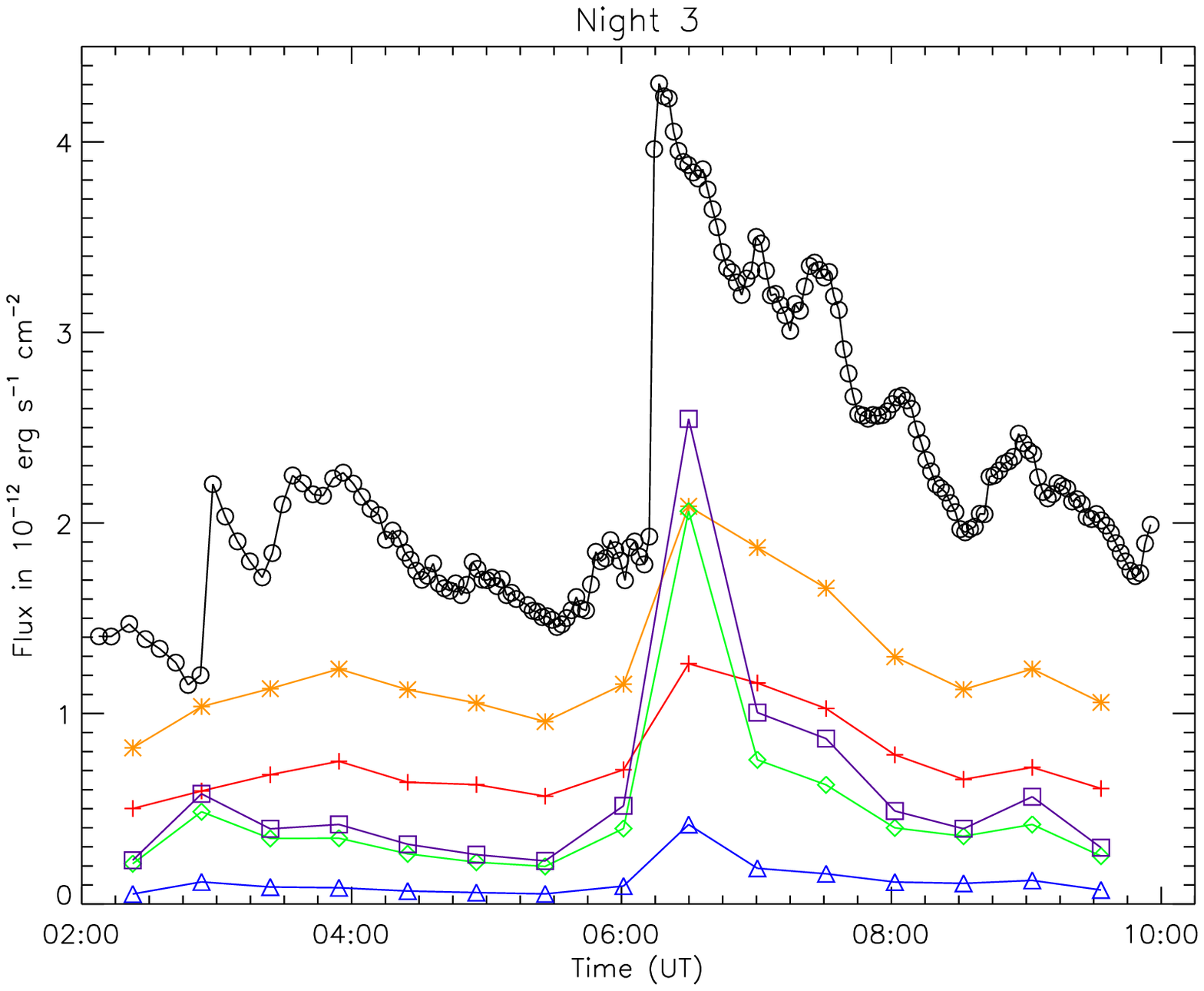}\vspace{-0.4mm}
\includegraphics[width=8cm]{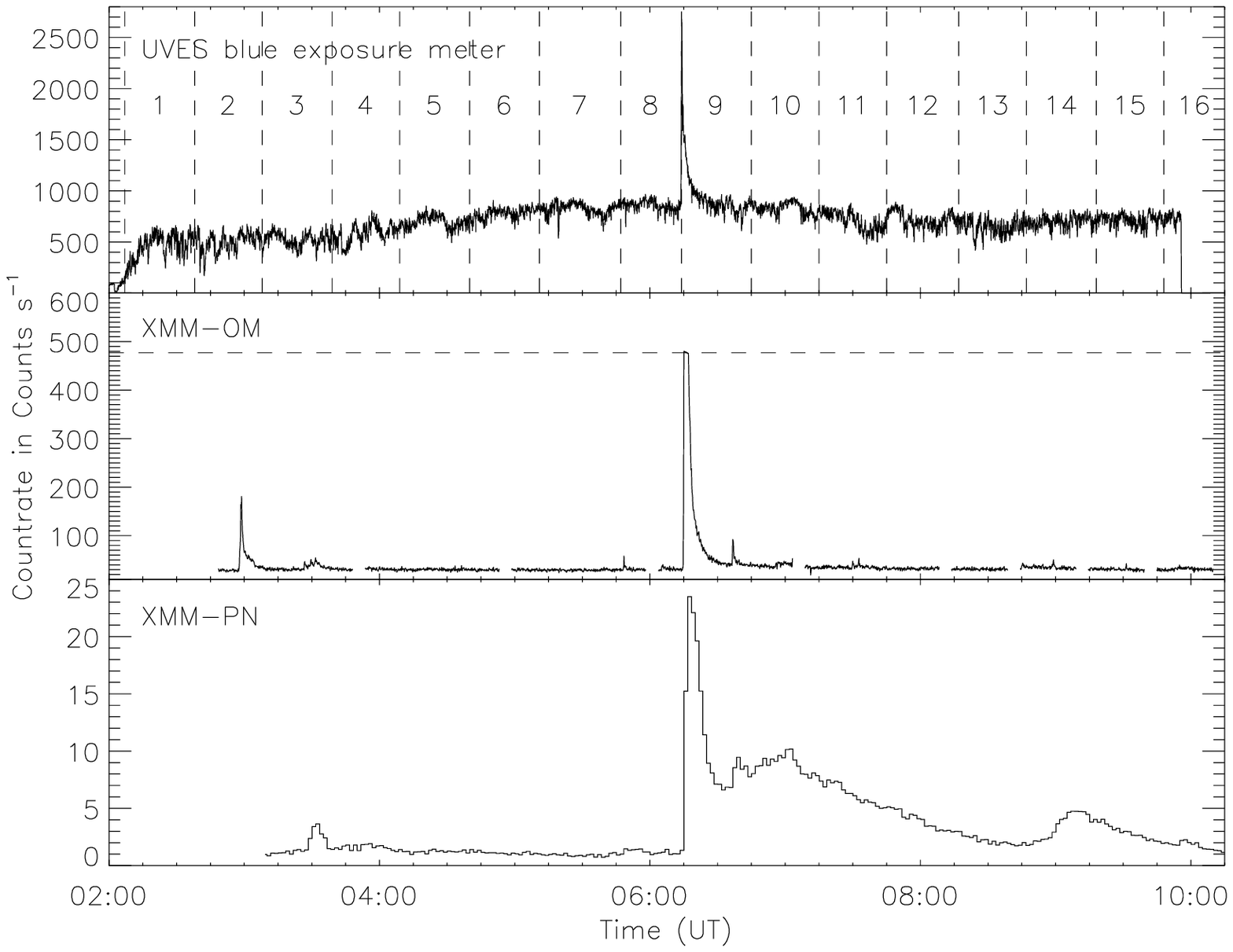}
\caption{\label{lightcurve3} As Fig.~\ref{lightcurve1}, but for night 3. The time intervals for the UVES blue arm spectra are also indicated 
for further reference. The dashed line in the OM panel marks the saturation limit that was reached in the flare peak.
 The figure legend is the same as in Fig. \ref{lightcurve1}.}
\end{center}
\end{figure}

In the optical (UVES exposure-meter and in the OM), flaring state and quiescent phases can easily be identified even for smaller flares. For the 
chromospheric emission line light curves and the X-ray light curves this is not so easy:
the lowest parts of the X-ray light curve -- as seen in Figs. \ref{lightcurve1} -- \ref{lightcurve3} --
show a typical count rate of $\approx 1$~cps in the PN detector,
comparable to the low count rate states found in the 2001 observations of Proxima Centauri (see \citet{Guedel_ProxCen_1}).
During these time intervals, the light curves also display some variability, i.\,e. do not remain constant. 
To be consistent with other authors, we nevertheless
use the  term "quiescence" and  define it for our subsequent spectral analysis as those times with
a PN count rate below $1.8$~cps.

In addition to the low-level variability seen during quiescence, small flares are observed at 3:30 and 6:00 UT in the X-ray data from night~1 
(see Fig. \ref{lightcurve1}). Similarly, in the X-ray 
data from night~2 (Fig. \ref{lightcurve2}), which have a slightly higher mean count rate, there is some low-level 
variation with small
flares at about 4:40, 6:15 and 8:40~UT. Most of these flares are also visible in the OM light curves, while the UVES 
exposure-meter does not show significant variability. Apart from these small flares, the X-ray light curves also show some 
plateau-like structures and slow rises and decreases  not seen in the OM light curve, which displays a 
constant quiescent level during these times.

Nevertheless, a comparison of the X-ray flux with several chromospheric
line fluxes (see Figs. \ref{lightcurve1} and \ref{lightcurve2}) shows quite  good agreement, especially
for the H$\alpha$ light curve, which can be studied with the much higher time-resolution of the red arm spectra.  
For instance, the first H$\alpha$ flare in night~1 at about 3:30~UT can also be
identified  in the X-ray light curve, although not in the optical (UVES) light curve. In addition, the flare in night~2
at 6:15 UT is found in X-ray, OM, and in the chromospheric emission lines, but is much more pronounced in the 
\ion{Ca}{ii} H \& K line than in the Balmer lines.
This shows that the chromospheric emission lines can  also be affected by small flares, when the continuum exhibits
no changes. Moreover, the chromospheric emission lines exhibit variability that is not seen in the X-ray band,
thus should be confined to the chromosphere; for instance, in night~2, during the first two hours, the H$\alpha$ and
other chromospheric emission lines display a flux decline, but the X-ray level is constant, and in night 1, 
the chromospheric lines show a peak at about 7:00~UT that is not noticed in the X-ray or continuum. This
behaviour during flares has been found before. \citet{Osten_EV_Lac_flare} found at radio, optical, and X-ray
wavelengths for the  flare star EV Lac flares with no counterparts at other wavelengths. In addition \citet{Hilton}
searched for flares in the SDSS and found 243 flaring spectra of 63 flare stars with only two having an enhanced continuum. 
Continuum enhancement occurs especially for the strongest flares (see e.\,g. \citet{Kowalski_2} and \citet{Hawley_Fisher}). 

\begin{figure}
\begin{center}
\includegraphics[width=8cm]{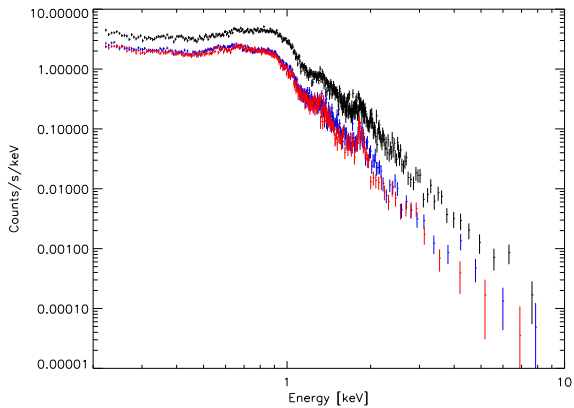}
\caption{\label{spectra} PN spectra from night 1 (bottom/red), night 2 (middle/blue), 
and night 3 (top/black).}
\end{center}
\end{figure}

We also analysed Proxima Centauri's integrated flux using PN spectra from the three individual 
{\em XMM-Newton} observations. The X-ray flux levels did not change significantly for the 
first and second exposure, whereas there is a significant change in the X-ray flux in the 
third exposure as can be seen in Fig.~\ref{spectra}; here the X-ray flux was determined using the Xspec best-fit spectral models (see Section 4) integrated in the energy range 0.2 to 10.0~keV. Specifically, the 
night 1 EPIC PN exposure results in an X-ray flux of $3.0\times 10^{-12}$
 erg\,cm$^{-2}$\,s$^{-1}$ ($L_{x}$ = $6.0\times 10^{26}$~erg\,s$^{-1}$), the  
exposure of night 2 corresponds to $3.2\times$ 10$^{-12}$~erg\,cm$^{-2}$\,s$^{-1}$ ($L_{x}$ = 
$6.4\times 10^{26}$~erg\,s$^{-1}$), and the data from night 3 give $6.3 \times 10^{-12}$
\,erg\,cm$^{-2}$\,s$^{-1}$ ($L_{x} = 1.2 \times 10^{27}$~erg\,s$^{-1}$). The higher value
for the third night is caused by the strong flare.  

During all three nights, the X-ray flux in the quiescent time intervals  is at a level of $2.6 \times 10^{-12}$\,erg\,cm$^{-2}$\,s$^{-1}$ ($L_{x} = 4.9 \times 10^{26}$~erg\,s$^{-1}$).
This is comparable to the lower boundary of the variable quiescent X-ray luminosity $4 - 16 \times 10^{26}\mathrm{~erg\,s^{-1}}$ found by \citet{Haisch}.

We can calculate the activity level of Proxima Centauri using the activity indicator log$L_X/L_{bol}$. 
With infrared H and K band magnitudes of $m_H =  4.835$ and $m_K =  4.384$ and using the bolometric corrections 
given by \cite{reid2001}, Proxima Centauri's bolometric luminosity is  $6\times 10^{30}$erg\,s$^{-1}$, i.e.,
slightly less than the $L_{bol}=6.7 \times 10^{30}$erg\,s$^{-1}$ found by \citet{Frogel}.
With an activity indicator of log$L_X/L_{bol}=-4$ in non-flaring time intervals, Proxima Centauri is a 
moderately active star.  

 \subsection{The large flare}


The X-ray light curve from night 3 (Fig.~\ref{lightcurve3}) shows a
large flare with a peak X-ray flux of $3.8 \times 10^{-11}$\,erg\,cm$^{-2}$\,s$^{-1}$ ($L_{x} = 7.2 \times 10^{27}$~erg\,s$^{-1}$), which is also evident  in the OM and optical data. 
The optical UVES exposure-meter and OM light curves 
resemble those of a typical
solar impulsive flare. The X-ray data and light curves of chromospheric emission lines
show a more complex behaviour with two broad secondary events peaking at about 6:45 and 9:10 UT in
the X-ray light curve. In the H$\alpha$-emission, we find an even more complicated behaviour with several 
sub-peaks in the decay phase, which are not seen in any of the other light curves but are roughly
associated with the two secondary peaks seen in X-rays. The
two secondary peaks can also be identified  in the other Balmer lines and the \ion{Ca}{ii} H and K lines
despite their lower time resolution. The secondary peaks are found in neither the optical UVES light curve,
nor  the OM light curve. 
A  similar event was described for Proxima Centauri by \citet{Guedel_ProxCen_1}.

The flare rise to the maximum flux took place on a timescale of 
$\sim$500 sec.  With a 10s binning for X-ray and optical data, we found that 
the optical peak as seen in the OM light curve precedes the X-ray peak by 
about two minutes. This can be explained in terms of the Neupert effect \citep{Neupert1968} 
known to occur in solar and a variety of stellar flares, where the time integral 
of emission due to particle acceleration, such as radio emission or blue 
continuum emission, resembles the rise of the flare light curve in 
soft X-rays. The explanation of this behaviour is that accelerated particles 
hit the dense chromosphere and that chromospheric material is heated and evaporates 
into the corona, where it accumulates and provides the emission measure of the 
subsequent soft X-ray emission. 

\begin{figure}
\begin{center}
\includegraphics[width=8cm,clip]{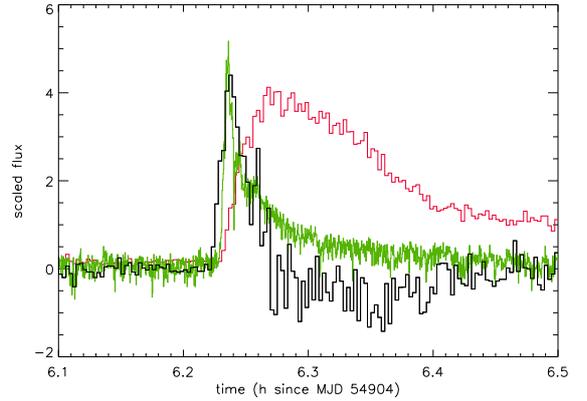}\vspace{-0.4mm}
\caption{\label{neupert} Illustration of the Neupert effect during the large flare of Proxima Centauri on night 3. 
Depicted are the combined EPIC X-ray light curve in red, its time derivative (smoothed by five bins) in black, 
and the optical UVES light curve in green.}
\end{center}
\end{figure}

This effect is also clearly seen in our data, because the time derivative of the combined 
EPIC (MOS1, MOS2 and PN) X-ray light curve during the flare rise 
(see Fig.~\ref{neupert}) matches the shape of the optical light curve; 
we note that we cannot use the OM light 
curve for this analysis, since
the OM data are saturated during the flare peak, but we use the light curve
that is extracted from the UVES's blue-arm photometer.
 
\subsection{Average magnetic field}\label{avmagfield}

The average magnetic field $Bf$ of Proxima Centauri is measured using our UVES red arm data by employing a method
introduced by \citet{2006ApJ...644..497R} utilising an absorption band
of molecular FeH close to 1 $\mu$m. We provide a brief overview of the
method here and refer to the paper by \citet{2006ApJ...644..497R} for a more detailed description of the 
procedure. The FeH band contains a large number of isolated lines, some of which
are highly sensitive to the Zeeman effect, while others are not. Since Land\'{e}-$g$ factors are unavailable, a direct
determination of the Zeeman effect is difficult, although there have been some
attempts to model FeH molecular lines theoretically (see \citet{ModZeeman2010}). 
We measure the magnetic field by comparing observed FeH lines to spectra of
stars with known magnetic fields. These template stars are GJ~1002
(M5.5, no measured magnetic field) and Gl~873 (M3.5, $Bf\sim3.9$~kG,
\citealt{2000ASPC..198..371J}). Before the spectra can be compared, the
template spectra have to be scaled to match the strengths of the FeH
absorption to the strength found in our spectra of Proxima Centauri. They also have to be
adjusted to match in terms of rotational velocity. The observed
spectrum is then modelled as a linear combination of the template spectra to
determine $Bf$, thus it is assumed that the Zeeman broadening is
linear in $Bf$ and that the magnetic field strength is
distributed similarly in all stars.

\begin{figure}
\begin{center}
\includegraphics[width=8cm]{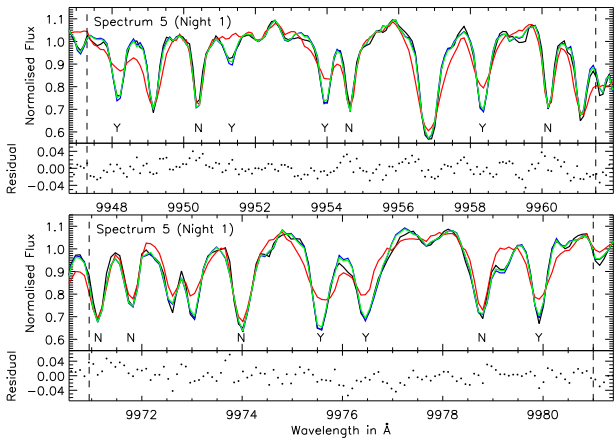}
\caption{\label{Bffitreg} Result of the fit for UVES red arm spectrum 5 from night 1 with 
$Bf=400\,\mathrm{G}\pm110\,\mathrm{G}$. \textit{Black}: slope corrected Proxima Centauri data; 
\textit{red}: scaled and broadened spectrum of magnetic reference Gl~873; \textit{blue}: 
scaled and broadened spectrum of non-magnetic reference GJ~1002; \textit{green}: best fit; 
\textit{Y}: magnetically sensitive lines; \textit{N}: magnetically insensitive lines; 
\textit{vertical dashed lines} border the wavelength regions used for the fit.}
\end{center}
\end{figure}

We measured the average magnetic field $Bf$ for all three nights by using a
$\chi^2$-minimisation to find the best-fit relation for the
interpolation between the template stars and the UVES red arm spectra of Proxima Centauri.
The fit was calculated in two wavelength regions, 9947.3\,\AA{}-9961.5\,\AA{}
and 9970.95~\AA{}-9981.0~\AA{}, containing 14 absorption lines of FeH with 7 being
sensitive and 7 being insensitive to the magnetic field, respectively. A typical example
of our fits is shown in Fig.~\ref{Bffitreg}. We compute a formal 1$\sigma$ uncertainty in $Bf$ by searching 
the range of $Bf$ for which $\chi^2 < \chi_{min}^{2}+1$ while 
varying the four other parameters. This results in an average uncertainty of $\overline{\Delta Bf}=110$~G. 
We note that it is only the statistical error that we use to compare individual \textit{differential} 
measurements because the magnetically insensitive lines remain constant; thus they provide an accurate calibration
particularly when intercomparing exposures. In addition to statistical errors, there are systematic errors and we estimate 
the true uncertainty in the absolute measurement of the magnetic field $Bf$ to be of the order of a 
few hundred gauss. Hence, to accurately assess any trend in the data the statistical error has to be considered,
while the systematic errors may offset the whole graph.

\begin{figure}
\begin{center}
\includegraphics[width=8cm]{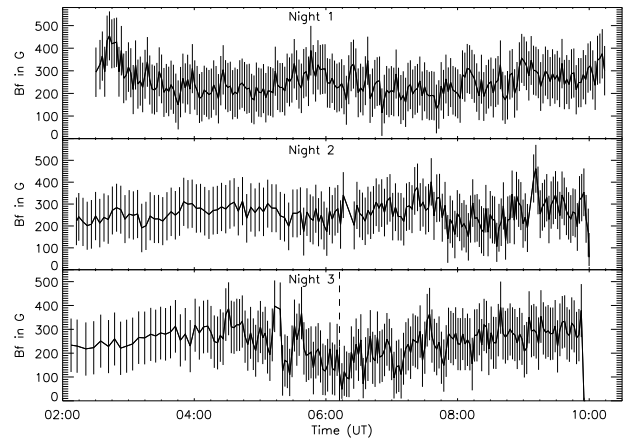}
\caption{\label{Bf} Timing behaviour of the average magnetic field $Bf$ for the
  three observed nights. The onset of the flare in the third night is marked by a vertical dashed line.}
\end{center}
\end{figure}

There is a relatively strong correlation between the seeing conditions and the derived values of $Bf$,
especially in the second and third night with Pearson's product-moment correlation coefficients $r$  of
$r_{N2}=0.84$ and $r_{N3}=0.64$. This was eliminated by
dividing the $Bf$ measurements by a polynomial describing the general
trend of the seeing taken from the ESO archive. The resulting timing behaviour of $Bf$ is shown
in Fig. \ref{Bf}. The field has a mean level of $\overline{Bf} = 250~\rm{G}$ with a statistical error of $ 60~\rm{G}$. 
\citet{ReinersBf} also measured the average magnetic field of Proxima Centauri
with the same method  using data taken in April 2004 and  found magnetic field values in the range 450 G $< Bf <$ 750 G. 
This change of $\sim$350\,G over five years might well be real, but is close to the limits                                                                          
of our uncertainty. 
Hence, it is unclear whether this difference is real or a 
consequence of the measuring inaccuracy.

As mentioned above, we can use the statistical error in the $\chi^2$-minimisation to compare 
individual measurements, because we are only interested in a differential comparison and 
not in the absolute values of $Bf$.
For the first and second nights, the magnetic field is consistent with a constant value within the error bars. For
the third night, there is a dip  in the magnetic field with its
minimum roughly at the time of the flare onset (Fig.~\ref{Bf}). In the context of
the reconnection model for flares, a weak magnetic field at flare onset is what one would
expect.  However, we note that the measured fields refer to the photosphere, while the
field annihilation is assumed to take place in the corona.  In addition,
the slow decay before the flare event is surprising for a catastrophic event such as magnetic 
reconnection, while after the flare the magnetic field recovers relatively quickly. 
 In summary, it remains unclear whether there is a physical
association of the magnetic field changes with the flare.

\section{Coronal properties of Proxima Centauri}

\subsection{X-ray spectral analysis}

Our X-ray observations of Proxima Centauri can be analysed to help us derive
 several time-variable properties of its corona such as temperatures, abundances, and 
densities, as we investigate in this section.

\subsubsection{Spectral fits and elemental abundances}\label{corabund}

\begin{figure*}
\begin{center}
\includegraphics[width=16cm]{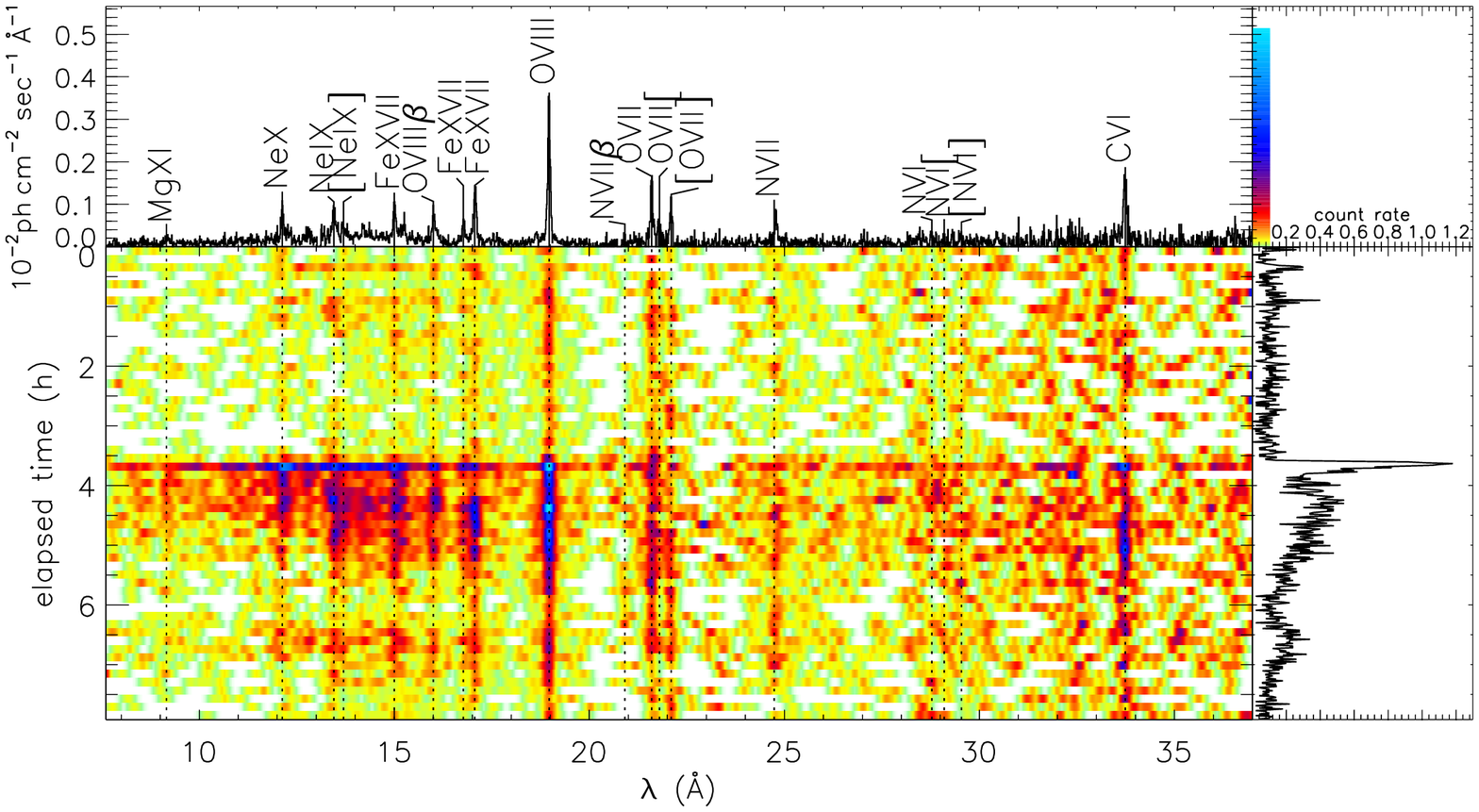}
\caption{\label{rgs} Time evolution of RGS spectra of Proxima  Centauri. 
Prominent emission lines are labelled in the mean spectrum shown in the upper part of the graph.
In the right panel, the associated light curve can be found.}
\end{center}
\end{figure*}

To visualise the spectral changes of Proxima Centauri associated with the large flare, 
we show the time evolution of the RGS spectra during night~3 in Fig.~\ref{rgs}. 
The mean spectrum is shown in the uppermost part of the diagram, while the time evolution in the
form of the light curve is depicted vertically at the right border. The mean spectrum 
resembles a typical RGS spectrum of M-dwarfs with \ion{O}{viii} 
Ly $\alpha$ being the strongest line and other prominent lines originating
from \ion{Ne}{ix}, \ion{Ne}{x}, \ion{Fe}{xvii}, \ion{O}{vii}, and \ion{C}{vi}. As expected, the diagram shows a 
brightening in the spectral lines with the flare onset. The higher plasma temperatures during the flare 
manifest themselves as a relative brightening of lines formed at higher coronal temperatures than the cooler 
spectral lines, for example \ion{O}{viii} vs. the \ion{O}{vii} triplet. However, pronounced changes in the 
density-sensitive ratio of the \ion{O}{vii} forbidden to intercombination line cannot be identified in this 
time-resolved plot for the flare and the pre-flare period; they are only evident in time-integrated measurements for the flaring and quiescent states (see section~\ref{densitiesne}).

There is  particular interest in studying elemental abundances during a
flare. Since flares transport chromospheric material into the
corona by chromospheric evaporation, coronal abundance patterns can 
temporarily change during flares. We determined the temperatures and abundances 
relative to solar values \citep{Grevesse} with an iterative procedure of 
global Xspec fits to EPIC and RGS spectra with VAPEC plasma models. In these fits, the
abundances and emission measure are inherently interdependent, hence we
made our inferences from relative changes in the fit parameters for the three observations. 
The errors we give for the fitted parameters were calculated with Xspec's {\it error} command.
The program Xspec calculates the confidence intervals for the desired parameter by fixing this parameter at a
specific value and varying the other free parameters until the best fit is reached. New fixed values for
the parameter of interest are chosen until the error margins that were requested in the {\it error}
command are covered. The errors in these 
plasma fits are correlated; in particular for the emission measure and elemental abundances, there is a strong 
interdependence (see below). Thus, in the calculation of Xspec's error margins for an individual parameter, 
all other non-frozen parameters are also allowed to vary, yielding effectively larger error margins than 
if one (incorrectly) assumes the errors to be uncorrelated.

When fitting the
EPIC spectra, we adopted the following method to obtain the
abundances: the abundances of oxygen, neon, iron, magnesium, and silicon 
were allowed to vary freely and independently, but were constrained
to be the same among all the VAPEC temperature components. 
The abundances of all other elements
were frozen to the solar value because they cannot be well-constrained  with EPIC 
spectra since they do not have prominent spectral lines at energies where EPIC has a large effective area.
When fitting EPIC spectra together with RGS spectra, carbon and nitrogen abundances also were allowed to vary. 
To determine the number of VAPEC components with different temperatures, we added new components 
one by one until the addition of
the next component did not improve the fit significantly. 

In the following sections we used the RGS spectra with their much higher wavelength resolution  
for spectral fitting simultaneously to the MOS spectra. We combined the spectra from 
the quiescent states of all three nights (for a definition of quiescence, see section 
\ref{quasi-quiescent}) and compared them to the spectra obtained from the large 
flare during the third night. To compare these two states, we defined a fixed temperature 
grid consisting of the values $0.14$, $0.4$, and $1.0$~keV ($1.6$, $4.6$, and $11.5$~MK). The 
resulting best-fit $\chi^{2}$ values 
of the three-temperature VAPEC models along with average individual elemental abundances for 
the quiescent state and the flaring state are listed in  Table \ref{3temp}.
The quiescent state is characterised by dominant plasma components at $0.14$ and $0.4$~keV. 
During the flare, there was a pronounced enhancement of the emission measure at $0.4$ and $1.0$~keV. 
The emission measure at low temperatures is more weakly constrained than for the quiescent state; within the errors, the low-temperature emission measure is consistent in both fits.

\subsubsection{The FIP effect}\label{fip}

\begin{figure}
\begin{center}
\includegraphics[width=8cm]{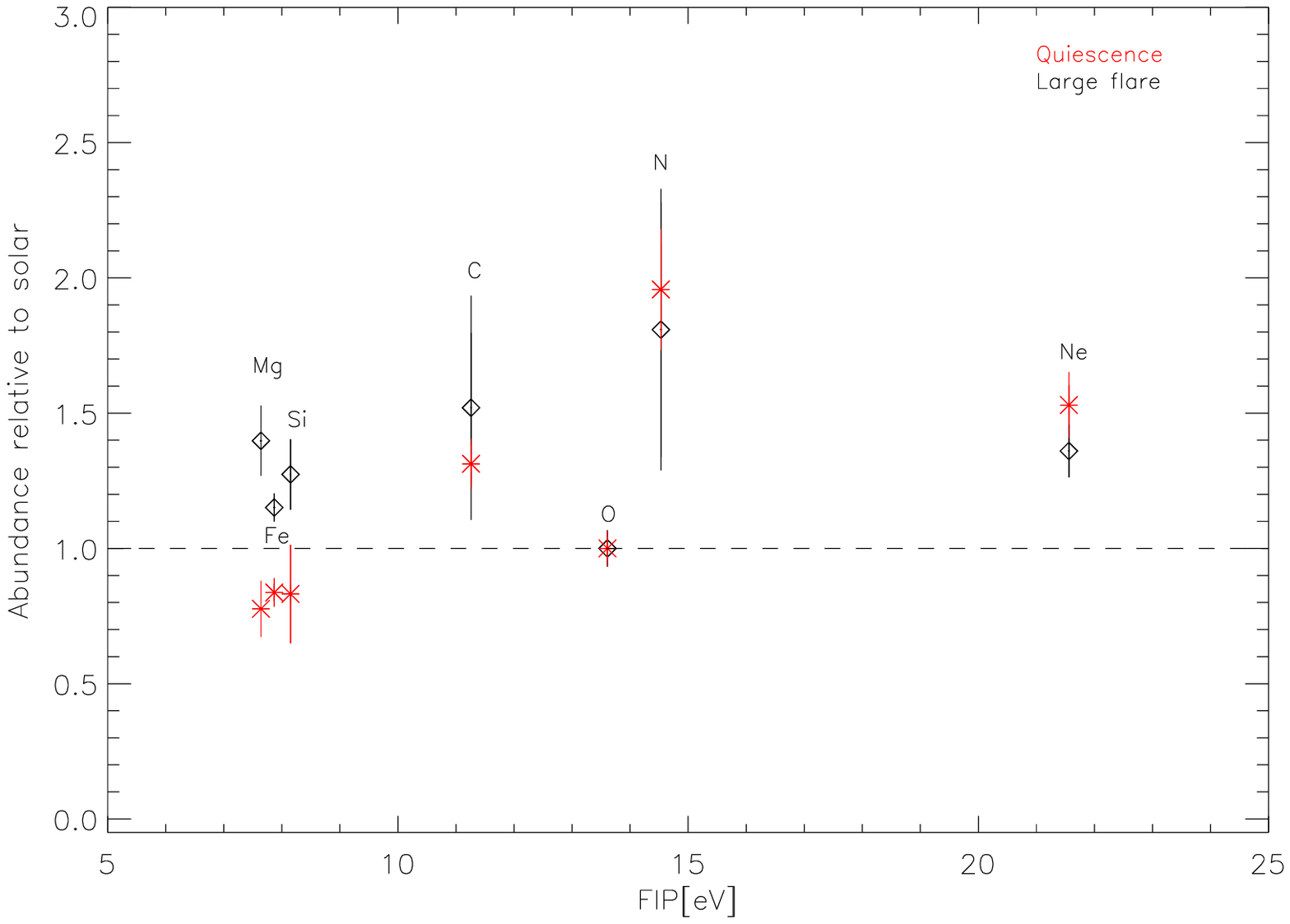}
\caption{\label{fip_graph} Normalised elemental abundances to oxygen abundance relative to solar photospheric values (\citep{Grevesse}) as a function of the first ionisation
potential (FIP) during quiescence (red) and flare (black). Dashed line indicates the solar photospheric abundance.}
\end{center}
\end{figure}

Similar to the Sun, inactive or low-activity-level stars
show a normal first ionisation potential (FIP) effect where elements with a low  FIP 
are enhanced in the corona compared to elements with a high FIP.
A reversed pattern - the inverse FIP effect - with enhanced high-FIP elements and 
depleted low-FIP elements is frequently found in stars of higher activity 
(e.g. \citet{Brinkman}, \citet{Audard}). 
A FIP effect is common among stars with 
low activity levels (log$L_X/L_{bol} < -4$). 
Owing to the activity level shown by Proxima Centauri, one would expect to observe the inverse FIP-effect. In the measurements for
the quiescent state, the inverse FIP-effect might be observed but owing to our measurement errors we cannot
make a strong claim here, which is illustrated in Fig.~\ref{fip_graph}. The actual abundance values can be found in Table \ref{3temp}.
During the flare, the abundances are slightly higher on average, and the iron 
and silicon abundances rise in relation to the high-FIP elements. This is in 
line with the picture that fresh chromospheric material is evaporated into the 
corona during the flare, which causes a change in elemental abundances. In 
previous X-ray observations of Proxima Centauri, a similar abundance distribution was found \citep{Guedel_ProxCen_2}.

\begin{table}
\caption{\label{3temp} Three-temperature fit to the X-ray spectra extracted from MOS1, MOS2, RGS1, and RGS2 data allowing individual 
elemental abundance and emission measure to vary on a fixed temperature grid. Errors given are $1\sigma$ errors.}
\begin{tabular}[htbp]{ccc}
\hline
\hline
Parameters & quiescence& large flare\\
\hline \\[-3mm]
$T_1$ (keV)	& \multicolumn{2}{c}{0.14}	\\

$EM_{1}$ ($10^{50}$ $cm^{-3}$) 	&$0.16_{-0.02}^{+0.02}$		&$0.17_{-0.07}^{+0.07}$	\\	
$T_2$ (keV)	& \multicolumn{2}{c}{0.4}	\\

$EM_{2}$ ($10^{50}$ $cm^{-3}$) 	&$0.35_{-0.02}^{+0.02}$		&$1.42_{-0.07}^{+0.08}$	\\	

$T_3$ (keV)	& \multicolumn{2}{c}{1.0}	\\

$EM_{3}$ ($10^{50}$ $cm^{-3}$) 	&$0.01_{-0.01}^{+0.01}$		&$1.10_{-0.03}^{+0.03}$	\\

C 		& 0.57$_{-0.03}^{+0.04}$	&0.66$_{-0.12}^{+0.18}$\\
N 		& 0.79$_{-0.07}^{+0.09}$	&0.73$_{-0.19}^{+0.21}$ \\
O 		& 0.45$_{-0.02}^{+0.03}$	&0.45$_{-0.03}^{+0.03}$\\
Ne 		& 0.63$_{-0.03}^{+0.05}$	&0.56$_{-0.04}^{+0.04}$\\
Mg 		& 0.30$_{-0.03}^{+0.04}$	&0.54$_{-0.05}^{+0.05}$\\
Si 		& 0.32$_{-0.07}^{+0.07}$	&0.49$_{-0.05}^{+0.05}$\\
Fe 		& 0.32$_{-0.02}^{+0.02}$	&0.44$_{-0.02}^{+0.02}$\\
red. $\chi^{2}$	&1.96				&1.877\\
D.O.F.		&784				&561\\
$\log L_X$ (0.2-10 keV)&26.70 			&27.50	\\
\hline
\end{tabular}
\end{table}

To  investigate in greater detail the abundance variations, we created spectra by dividing the flare data into several 
time intervals with the first four
lasting 120s, followed by three intervals of 180s, 240s, and 300s and 
the remaining data being divided into spectral intervals of 600s each. 
The first spectrum covers the flare rise and the following
time intervals cover the different phases of the decay. These spectra are 
fitted with combinations of APEC models,
using two temperature components and assuming fixed elemental abundances for the 
flare plasma but leaving the iron abundance 
as a free parameter. To obtain the plasma properties, we use the 
total emission measure EM, i.e. the sum
of the emission measures of each temperature component 

\[
EM ={\sum_{i}} EM_{i}
\]
and the flare temperature T is defined as an emission-measure weighted 
sum of the temperatures
from each flare component  \[
T ={\sum_{i}} \frac {T_{i} \times{ EM_{i}}}{EM}.
\]


We infer that the iron abundance
increases from a level of 0.30 $\pm$ 0.03
directly before the onset of the flare to a maximum
value of $0.58_{-0.12}^{+0.15}$, when the flare peak is reached (see Table \ref{model}). As oxygen is another element that produces very strong emission lines in our X-ray spectra, 
we also investigated possible oxygen abundance changes during the flare in the same fashion, 
but found no evidence of a similar timing behaviour. This indicates that fresh plasma material is 
evaporated from the photosphere and chromosphere that has a different composition with a higher iron abundance 
than the quiescent coronal plasma. This scenario fits very well with the measured H$\alpha$ asymmetries
that indicate movements from the chromosphere in the direction of the corona (see section \ref{asymmetries}).
As can be seen from line shift measurements in Table \ref{Hashifts}, material is evaporated with a 
velocity of 44.3 $\rm km\,s^{-1}$ in the spectrum lasting
from 6:12 to 6:14 UT. If one makes a conservative estimate that the material is accelerated only at the end of
the exposure time, it can travel for about 90 seconds or about 4000 km before the measurement no. 1 of 
Table \ref{model} starts (6:16:00 UT), which would place the material within the  corona. For the second measurement of 
H$\alpha$ blue-shifts, the situation is not as clear because the blue-shifts persist for a longer time overlapping with
measurement no. 3 (starting at 6:20:00 UT) of higher coronal iron abundance. If most of the material is 
evaporated at the beginning of
the blue-shift, the material moves for about 210 seconds or about 2200 km before the X-ray measurement no. 3 starts,
again putting the material into the  corona.

\begin{table*}
\caption{\label{model} Model parameters for the flare spectra with variable Fe abundances.}
\begin{tabular}[htbp]{cccccccccc}
\hline
\hline
Spect.& KT$_{1}$ & EM$_{1}$ & KT$_{2}$ & EM$_{2}$ & Fe & red. $\chi^{2}$ & d.o.f & T & EM \\
No.& [keV] & [$10^{50}cm^{-3}$] & [keV] & [$10^{50}cm^{-3}$] & & & &MK &  [$10^{50}cm^{-3}$]\\
\hline \\[-3mm]
1 & $0.75_{-0.06}^{+0.05}$ & 2.38$_{-0.63}^{+0.91}$& 1.90$_{-0.26}^{+0.28}$ &4.90 $_{-0.72}^{+0.58}$ & 0.58$_{-0.12}^{+0.15}$ &1.15 &119 & 17.71$_{-1.95}^{+1.65}$ & 7.28$_{-1.36}^{+1.49}$ \\
2 & $0.69_{-0.09}^{+0.07}$ & 3.23$_{-0.85}^{+1.12}$ & 1.58$_{-0.17}^{+0.19}$ & 6.39$_{-0.98}^{+0.81}$ & 0.38$_{-0.07}^{+0.12}$ & 1.10 &148  & $14.88_{-1.61}^{+1.44}$ & $9.62_{-1.83}^{+1.93}$ \\
3 & $0.61_{-0.03}^{+0.03}$ & $2.85_{-0.44}^{+0.53}$ & 1.60$_{-0.12}^{+0.11}$ & 4.98$_{-0.42}^{+0.40}$ & 0.54$_{-0.09}^{+0.11}$ &1.19 &159 & $14.40_{-0.91}^{+0.82}$ & $7.83_{-0.86}^{+0.94}$ \\
4 & $0.63_{-0.09}^{+0.03}$ & $3.61_{-1.87}^{+1.47}$ & 1.26$_{-0.15}^{+0.27}$ & 2.85$_{-0.92}^{+1.71}$ & 0.35$_{-0.09}^{+0.11}$ & 0.95 &93 & $10.60_{-1.32}^{+1.89}$ & $6.47_{-2.80}^{+3.19}$ \\
5 & $0.58_{-0.12}^{+0.07}$ & $2.41_{-0.54}^{+0.59}$ & 1.22$_{-0.24}^{+0.48}$ & 1.91$_{-0.62}^{+0.48}$ & 0.33$_{-0.06}^{+0.07}$ & 1.07 &113 & $10.05_{-2.17}^{+2.95}$ & $4.32_{-1.16}^{+1.07}$ \\
6 & $0.46_{-0.10}^{+0.13}$ & $1.80_{-0.36}^{+0.75}$ & 0.97$_{-0.13}^{+0.14}$ & 1.71$_{-0.75}^{+0.38}$ & 0.24$_{-0.07}^{+0.11}$ & 0.92 &96 & $8.26_{-1.41}^{+1.57}$ & $3.52_{-1.12}^{+1.13}$ \\
7 & $0.38_{-0.06}^{+0.09}$ & $1.16_{-0.18}^{+0.17}$ & 1.06$_{-0.07}^{+0.11}$ & 1.67$_{-0.11}^{+0.22}$ & 0.37$_{-0.10}^{+0.20}$ & 1.20 &106 & $9.11_{-0.75}^{+1.20}$ & $2.84_{-0.30}^{+0.40}$ \\
8 & $0.79_{-0.03}^{+0.02}$ & $2.72_{-0.46}^{+0.34}$ & 1.58$_{-0.46}^{+0.89}$ & 0.79$_{-0.39}^{+0.45}$ & 0.21$_{-0.03}^{+0.05}$ &1.00 &135 & $11.28_{-2.69}^{+6.05}$ & $3.52_{-0.85}^{+0.79}$ \\
9 & $0.64_{-0.03}^{+0.02}$ & $1.71_{-0.44}^{+0.43}$ & 1.01$_{-0.14}^{+0.12}$ & 1.82$_{-0.35}^{+0.38}$ & 0.26$_{-0.03}^{+0.04}$ &1.03 &229 & $9.64_{-0.97}^{+0.82}$ & $3.53_{-0.79}^{+0.82}$ \\
10& $0.62_{-0.12}^{+0.02}$ & $2.86_{-0.33}^{+0.39}$ & 1.22$_{-0.27}^{+0.42}$ & 0.92$_{-0.32}^{+0.28}$ & 0.30$_{-0.03}^{+0.03}$ &1.03 &222 & $8.88_{-2.32}^{+2.27}$ & $3.78_{-0.65}^{+0.67}$ \\
11& $0.46_{-0.09}^{+0.03}$ & $1.88_{-0.69}^{+0.74}$ & 0.76$_{-0.06}^{+0.13}$ & 2.00$_{-0.87}^{+0.72}$ & 0.27$_{-0.03}^{+0.05}$ & 1.00 &217& $7.17_{-0.91}^{+1.22}$ & $3.89_{-1.57}^{+1.47}$ \\
12& $0.31_{-0.04}^{+0.03}$ & $1.52_{-0.32}^{+0.29}$ & 0.72$_{-0.04}^{+0.05}$ & 1.72$_{-0.29}^{+0.34}$ & 0.36$_{-0.06}^{+0.07}$ & 0.92 &193& $6.21_{-0.51}^{+0.52}$ & $3.24_{-0.62}^{+0.64}$ \\
13& $0.24_{-0.03}^{+0.05}$ & $1.25_{-0.34}^{+0.45}$ & 0.64$_{-0.03}^{+0.05}$ & 1.70$_{-0.37}^{+0.35}$ & 0.38$_{-0.05}^{+0.07}$ & 0.81 &184 & $5.53_{-0.43}^{+0.62}$ & $2.95_{-0.72}^{+0.80}$ \\
14& $0.25_{-0.04}^{+0.05}$ & $1.04_{-0.35}^{+0.50}$ & $0.59_{-0.03}^{+0.06}$ & $1.42_{-0.46}^{+0.38}$ & $0.38_{-0.07}^{+0.09}$ &0.97 &162 & $5.23_{-0.47}^{+0.73}$ & $2.47_{-0.82}^{+0.89}$ \\
15& $0.22_{-0.05}^{+0.03}$ & $0.88_{-0.40}^{+0.29}$ & 0.57$_{-0.07}^{+0.04}$ & 1.36$_{-0.29}^{+0.54}$ & 0.34$_{-0.07}^{+0.08}$ &1.14 &142  & $5.05_{-0.74}^{+0.46}$ & $2.25_{-0.70}^{+0.83}$ \\
16& $0.26_{-0.01}^{+0.03}$ & $1.67_{-0.12}^{+0.12}$ & 1.78$_{-0.09}^{+0.15}$ & 0.40$_{-0.14}^{+0.20}$ & 0.35$_{-0.17}^{+0.21}$ &0.75 &134 & $4.28_{-0.69}^{+1.30}$ & $2.08_{-0.27}^{+0.32}$ \\
17& $0.24_{-0.01}^{+0.01}$ & $1.31_{-0.11}^{+0.10}$ & 0.78$_{-0.28}^{+0.19}$ & 0.25$_{-0.09}^{+0.44}$ & 0.37$_{-0.03}^{+0.05}$ & 1.11 &103& $3.82_{-1.61}^{+1.83}$ & $1.57_{-0.21}^{+0.55}$ \\
18& $0.24_{-0.02}^{+0.02}$ & $1.15_{-0.09}^{+0.14}$ & 0.70$_{-0.09}^{+0.13}$ & 0.25$_{-0.14}^{+0.11}$ & 0.36$_{-0.17}^{+0.37}$ & 1.02 &91 & $3.78_{-0.78}^{+0.83}$ & $1.40_{-0.24}^{+0.25}$ \\
19& $0.31_{-0.04}^{+0.09}$ & $1.15_{-0.30}^{+0.29}$ & 0.68$_{-0.22}^{+0.35}$ & 0.21$_{-0.19}^{+0.26}$ & 0.29$_{-0.07}^{+0.07}$ & 0.95 &74 & $4.29_{-1.39}^{+2.54}$ & $1.37_{-0.50}^{+0.56}$ \\
20& $0.22_{-0.02}^{+0.02}$ & $0.79_{-0.09}^{+0.08}$ & 0.69$_{-0.09}^{+0.14}$ & 0.15$_{-0.06}^{+0.09}$ & 0.20$_{-0.11}^{+0.07}$ & 1.24 &62 & $3.48_{-0.61}^{+1.01}$ & $0.95_{-0.16}^{+0.18}$ \\
21& $0.13_{-0.01}^{+0.03}$ & $0.72_{-0.19}^{+0.23}$ & 0.47$_{-0.06}^{+0.14}$ & 0.54$_{-0.30}^{+0.19}$ & 0.24$_{-0.08}^{+0.05}$ &0.98 &60 & $3.25_{-0.54}^{+0.97}$ & $1.26_{-0.50}^{+0.43}$ \\
22& $0.23_{-0.03}^{+0.02}$ & $0.70_{-0.13}^{+0.19}$ & 0.76$_{-0.10}^{+0.10}$ & 0.45$_{-0.13}^{+0.15}$ & 0.32$_{-0.12}^{+0.16}$ & 1.03 &77  & $5.14_{-0.84}^{+0.73}$ & $1.15_{-0.26}^{+0.34}$ \\
23& $0.40_{-0.06}^{+0.08}$ & $1.05_{-0.13}^{+0.21}$ & 0.91$_{-0.11}^{+0.07}$ & 1.00$_{-0.17}^{+0.25}$ & 0.33$_{-0.05}^{+0.08}$ &1.09 &135 & $7.57_{-1.05}^{+0.94}$ & $2.05_{-0.31}^{+0.46}$ \\
24& $0.26_{-0.04}^{+0.07}$ & $0.76_{-0.23}^{+0.31}$ & 0.72$_{-0.05}^{+0.08}$ & 1.22$_{-0.28}^{+0.25}$ & 0.27$_{-0.06}^{+0.08}$ & 0.84 &129& $6.38_{-0.59}^{+0.90}$ & $1.98_{-0.52}^{+0.57}$ \\
25& $0.19_{-0.03}^{+0.03}$ & $0.66_{-0.14}^{+0.13}$ & 0.63$_{-0.04}^{+0.04}$ & 0.90$_{-0.17}^{+0.19}$ & 0.32$_{-0.07}^{+0.09}$ & 1.14 &105 & $5.16_{-0.44}^{+0.48}$ & $1.56_{-0.31}^{+0.32}$ \\
26& $0.29_{-0.05}^{+0.06}$ & $0.94_{-0.22}^{+0.22}$ & 0.78$_{-0.14}^{+0.11}$ & 0.38$_{-0.11}^{+0.21}$ & 0.32$_{-0.15}^{+0.28}$ & 1.22 &85 & $5.09_{-1.00}^{+1.02}$ & $1.33_{-0.34}^{+0.44}$ \\
27& $0.26_{-0.03}^{+0.06}$ & $0.72_{-0.18}^{+0.17}$ & 0.68$_{-0.10}^{+0.12}$ & 0.32$_{-0.10}^{+0.18}$ & 0.32$_{-0.11}^{+0.18}$ & 1.02 &69 & $4.56_{-0.72}^{+1.09}$ & $1.04_{-0.29}^{+0.36}$ \\
28& $0.23_{-0.11}^{+0.02}$ & $0.65_{-0.16}^{+0.09}$ & 0.78$_{-0.16}^{+0.11}$ & 0.25$_{-0.08}^{+0.21}$ & 0.25$_{-0.20}^{+0.11}$ & 1.14 &66 & $4.49_{-1.53}^{+1.03}$ & $0.91_{-0.25}^{+0.31}$ \\
\hline
\end{tabular}
\end{table*}

\subsubsection{Coronal densities}\label{densitiesne}

Electron densities can be inferred from line triplets of He-like ions. The RGS energy range contains 
density-sensitive He-like triplets of \ion{N}{vi}, \ion{O}{vii}, \ion{Ne}{ix}, \ion{Mg}{xi}, and \ion{Si}{xiii}. 
These He-like triplets show in increasing order of wavelength a resonance (r), intercombination (i), and a 
forbidden line (f). If the electron collision rate is sufficiently high, ions in the upper level of the forbidden 
transition,$ 1s2s  {^ 3}S_{1}$, do not return to the ground level, $1s^{2}  {^ 1}S_{0}$, but  the ions are  instead
collisionally excited to the upper level of the intercombination transitions, $1s2p  {^ 3}P_{1,2}$, from where 
they decay radiatively to the ground state. Thus, the ratio of the f to i fluxes is sensitive to 
density (\citet{Gabriel_Jordan}).  

\begin{figure}
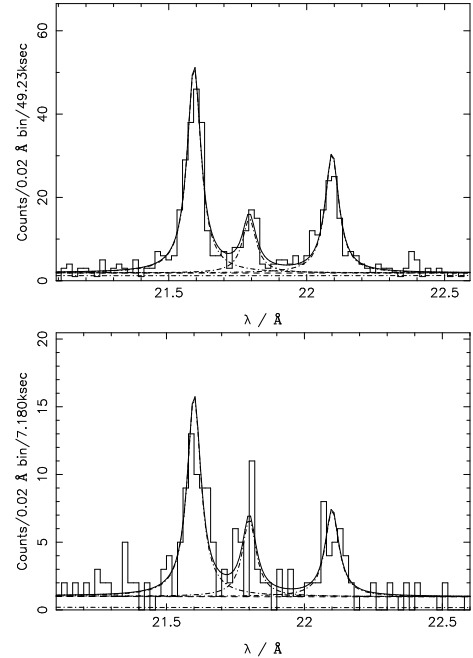

\begin{center}
\includegraphics[angle=270,width=6cm]{17447F10.ps}
\includegraphics[angle=270,width=6cm]{17447F11.ps}
\caption{\label{density} Density-sensitive line triplet of \ion{O}{vii} 
(resonance, intercombination, and forbidden line for increasing wavelength).
 The upper panel shows the best fit to cumulated RGS1 data during quiescence, while the
lower panel shows the RGS1 triplet data during the large flare.}
\end{center}
\end{figure}

The He-like triplet of \ion{O}{vii} is strong enough in our observations to be used to obtain 
characteristic electron densities in the source region. We estimate the coronal plasma densities using the density-sensitive 
ratio of the forbidden to intercombination line  of the \ion{O}{vii} triplet in the quiescent
and the flaring state. The measured $f/i$ flux ratios are $2.17 \pm 0.61$ in quiescence and $1.23 \pm 0.46$ 
during the flare in the third night (see Fig.~\ref{density}, Table~\ref{oviitable}). The errors in these line ratios are rather large and
are caused by the weak intercombination line. Formally, the errors overlap, hence one could argue that the change in density is insignificant.
However, the ratio $(f+i)/r$ is practically constant during quiescence ($0.84\pm 0.17$) and the flare ($0.90\pm 0.28$). 
This indicates that physically  X-ray photons are only shifted from the forbidden to the intercombination line at high densities. 
Hence, the constant $(f+i)/r$ ratio suggests that the change in the $f/i$ ratio is actually density-related and makes it less probable that it is a statistical fluctuation.

\begin{table}
\caption{\label{oviitable} X-ray counts in the \ion{O}{vii} triplet during quiescence (total integration time of 49~ks) and during the flare (total integration time of 7~ks).}
\begin{tabular}[htbp]{lcc}
\hline
\hline
line & counts (quiescence) & counts (flare)\\
\hline \\[-3mm]
resonance (21.6 \AA) 		& $225\pm 17$ 		& $96\pm 12$\\
intercombination (21.8 \AA) 	& $60\pm 11$		& $39\pm 8$ \\
forbidden (22.1 \AA) 		& $130\pm 13$		& $48\pm 8$\\
\hline
\end{tabular}
\end{table}

To convert the measured $f/i$ ratios into coronal plasma
densities, we approximated the flux ratio by $f/i = R_{o}/(1+n_{e}/ N_{c})$, where $R_{o}$ is 
the low density limit and $N_{c}$ is the critical density. We adopted the values from \citet{Pradhan} 
of 3.95 and 3.1 $\times$ 10$^{10}$~cm$^{-3}$, respectively. We find electron 
densities of $n_{e} = 2.5\pm 0.7 \times 10^{10}$~cm$^{-3}$ in quiescence 
and $n_{e} = 6.9\pm 2.6 \times 10^{10}$~cm$^{-3}$ during the third-night flaring period.
This suggests an increase in
plasma density during the flare, as one expects from the standard flare model.

\subsection{Interpretation: loop properties}\label{loopEMT}

Emission measure - temperature (EM-T) diagrams are useful for 
estimating physical quantities that are not directly observable. Since there 
is no direct information about the morphology of the involved coronal structures,
we estimate the size of the spatially unresolved stellar coronal flaring region 
from the light curve, time-resolved temperature (T), and emission measure (EM) 
values obtained during the decay of the flaring loop. 
The decay time of an X-ray flare is directly related to the length of the flaring loop. 
However, if there is a significant amount
 of heating present during the decay of the flare, it can prolong the decay and lead 
to an overestimated loop length.
We assumed that the flare occurs inside a single coronal loop with a constant cross-section 
that is anchored in the photosphere. The plasma is confined to the loop, and the 
decay begins after the loop has reached a quasi-steady state, where the energy and 
plasma flow are negligible. According to this assumption, the decay time of the X-ray 
emission roughly scales with the plasma cooling time, which 
in turn scales with the length of the loop structure. 
Therefore, the longer the decay takes, the larger  the loop structure becomes.
\cite{Serio} derived a thermodynamic timescale for pure cooling of 
flaring plasma confined to a single flaring coronal loop. \cite{Reale} 
derived an empirical formula  to determine the loop length considering the
effect of sustained heating during the flare decay that uses the trajectory of the flare in the EM-T diagram, viz.
\newline
 $L$  = $\frac{\tau_{LC} \sqrt{T_{0}}}{\alpha F(\zeta)}$ or $L_{9}= \frac{\tau_{LC}
 \sqrt{T_{0,7}}}{120 F(\zeta)}$  $\zeta_{min}<\zeta\leq\zeta_{max}$,
\newline

\noindent where $\tau_{LC}$ is the decay time derived from the light curve and  $\alpha=3.7 \times
 10^{-4}$ cm$^{-1}$s\,K$^{1/2}$. The observed maximum temperature must be corrected to
 $T_{0} (T-{0,7}) = \xi T_{obs}^{\eta}$ (in units of $10^{7}$ K), where $T_{obs}$ 
is the maximum best-fit temperature derived from spectral fitting to the data.  
The unit-less correction factor is $F(\zeta)= \frac{c_{a}}{\zeta - 
\zeta_{a}} + q_{a}$, where $\zeta$ is the slope of EM-T diagram. The coefficients
 $\xi$, $\eta$, $c_{a}$, $\zeta_{a}$, and $q_{a}$ depend on the energy response of the 
instrument. According to \citet{2007}, for XMM/EPIC the values are $\xi = 0.13$,
 $\eta = 1.16$, $c_{a} = 0.51$, $\zeta_{a} = 0.35$, and $q_{a} = 1.35$.

\begin{figure}
\begin{center}
\includegraphics[width=8cm]{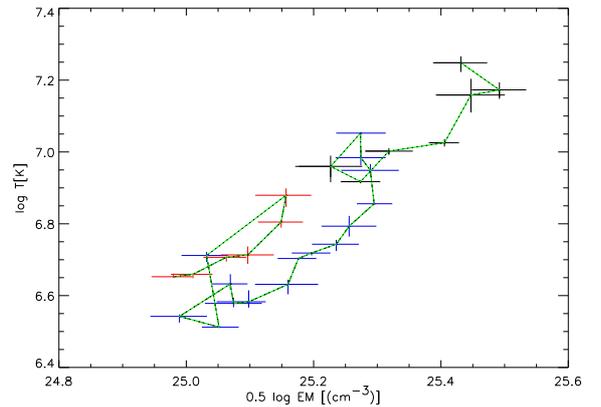}
\caption{\label{em_t} Flare evolution in density-temperature. 
The green line represents the evolution of flare decay (black) and  secondary events (blue and red)}
\end{center}
\end{figure}

We show the EM-T diagram of the flare on Proxima Centauri including the two secondary flares in Fig. \ref{em_t}.
For the part of the flare evolution before the secondary events, the slope measured from the 
logT - 0.5logEM  diagram in Fig. \ref{em_t}  is $\zeta = 1.07\pm0.29$. Given 
$T_{peak}$[MK] = $33.26_{-11.93}^{+7.73}$, we determined the loop half length to be $L=8.55_{-2.86}^{+3.81}$
 $\times 10^{9}$cm. Assuming that the volume $V \propto L^{3}$, we computed the volume of
 the flaring region $V_{flare} \approx 6.25_{-0.35}^{+2.12} \times 10^{29}$\,cm$^3$. We also computed the
loop-footpoint area $V/2L$ and an area filling-factor of the flare of three percent. 
We also note that in addition to  
 this flare decay, there are two other secondary events. From the EM-T diagram for these events, 
the slopes measured from these secondary events are similar to the slope obtained for the first flare event. 
Thus, the derived loop lengths are of the same order of magnitude within the errors.


\subsection{Optical \ion{Fe}{xiii} emission}

We also investigated the \ion{Fe}{xiii} forbidden coronal line at 3388.1~\AA\, in
our spectra of Proxima Centauri. The stellar spectrum had previously been searched for this
line before but no clear conclusion had been drawn \citep{FeXIII}. In the new UVES data, there is indeed evidence 
of the forbidden line during and after flare activity for the first and the
third night. 

\begin{figure}
\begin{center}
\includegraphics[width=8cm]{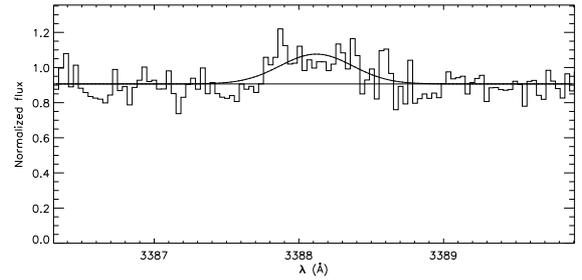}
\caption{\label{Fexiii} Example of the fit to the averaged difference spectrum of the
first night for the after-flare state at the end of the night for the \ion{Fe}{xiii} line.
The broad Gaussian
is the line best fit obtained with CORA. }
\end{center}
\end{figure}

To isolate the forbidden coronal line emission at 3388.1~\AA,  we averaged at least three normalised spectra for the quiescent state, the
flare state, and after-flare state. To emphasise the chromospheric and coronal 
changes between
the averaged spectra, we subtracted the quiescent state from the flaring state
and the post-flare state (and for testing purposes also from different
quiescent states of the same night). These difference spectra of the different
quiescent states usually contain only noise as expected. 
However, the differences between the spectra of flare
states and quiescent states in the first and third nights show a weak broad excess
at the position of the \ion{Fe}{xiii} line beginning at about UT 6:00 for the first night. 
This line persists for the post-flare
state. In the third night, the line can also be found directly before the large flare starting at UT 4:40, while no 
evidence of \ion{Fe}{xiii} emission could be found for the second night. 
An example of the line can be found in Fig. \ref{Fexiii}, the fit parameters 
obtained with CORA can be found in Table \ref{Fexiiitab}. The derived fluxes have an error of about a factor
of two because of the uncertainties in the flux calibration. For the first night, we used
the spectra no. 3--7 as quiescent state, for the second night the spectra no. 
4--6, and for the third night the spectra no. 1--5. The half width $\sigma$ of the line is 
much larger than the typical measured half width of the chromospheric lines  of 0.04 \AA\, (see Table \ref{linetable}).
Interpreting the line width as pure thermal Doppler broadening, we expect a half width of 0.04 \AA, for a 
typical chromospheric temperature of about 10\,000 K and a half width of about 0.17 \AA\, for the coronal \ion{Fe}{xiii} line
with a peak formation temperature of about 1.6 MK. The
measured half widths are larger for the forbidden \ion{Fe}{xiii} line suggesting that there are additional line-broadening mechanisms;
however, it must be kept in mind that the measurement of the line
width is sensitive to the determination of the background, 
thus the given (formal) errors are probably underestimates.

 The order of magnitude of the derived 
\ion{Fe}{xiii} flux is consistent with our estimate derived from Proxima~Centauri's coronal X-ray 
spectrum during the flare. The \ion{O}{vii} triplet is formed at a similar temperature as 
\ion{Fe}{xiii} ($\log T = 6.3$ and $6.2$, respectively). We calculate the flux in the \ion{O}{vii} 
triplet during the flare from a CORA fit to the RGS1 spectrum to be 
$F_{\ion{O}{vii}} = 4.32\times 10^{-13}$\,erg\,s$^{-1}$\,cm$^{-2}$. The coronal abundance ratio of 
iron to oxygen is $0.67$ during the flare (see Table~\ref{3temp}), and from CHIANTI we calculate 
the ratio of the line emissivities of \ion{Fe}{xiii} to the \ion{O}{vii} triplet to be of the order of 
0.25 in the temperature range $\log T = 6.2-6.3$. From this, we derive an order-of-magnitude 
estimate of $F_{\ion{Fe}{xiii}}\approx  10^{-15}$\,erg\,s$^{-1}$\,cm$^{-2}$, which is a little bit lower
but given the uncertainties compatible with 
the flux computed from the optical spectra in Table~\ref{Fexiiitab}.

The observation of the \ion{Fe}{xiii} line during the larger flare in the third night
seems to be counter-intuitive at first glance, since a flare should heat the corona
and \ion{Fe}{xiii} has its peak ionisation equilibrium at $T=1.6$~MK. 
However, during the flare the emission measure for the low temperature component at 1.6 MK is slightly higher than for quiecent state (or, 
given the errors, comparable to the emission measure during quiescence) and the iron abundance is
similary higher, which is consistent with the forbidden \ion{Fe}{xiii} line being seen during the flare. Moreover,
for another M5.5 star, LHS 2076, the line was also observed only during a flare \citep{FeXIII}.

\begin{table}
\caption{\label{Fexiiitab}Fit parameters of the \ion{Fe}{xiii} line.}  
\begin{minipage}{8cm}
\begin{tabular}[htbp]{cccc}
\hline
\hline
averaged & central wavelength & half width & line flux\footnote{Flux errors are about a factor of two.} \\
spectra no. & [\AA] & [\AA] & [erg\,s$^{-1}$\,cm$^{-2}$]\\
\hline
\multicolumn{4}{l}{1st night}\\
\hline \\[-3mm]
13--17 & 3388.02 $\pm$ 1.32 & 0.25 $\pm$ 0.02 & 6.6$\cdot10^{-15}$ \\
18--22 & 3388.07 $\pm$ 1.32 & 0.22 $\pm$ 0.02 & 6.6$\cdot10^{-15}$ \\
23--26 & 3388.10 $\pm$ 1.32 & 0.25 $\pm$ 0.02 & 7.2$\cdot10^{-15}$  \\
\hline
\multicolumn{4}{l}{3rd night}\\
\hline\\[-3mm]
6--8 & 3388.13 $\pm$ 1.32 & 0.27 $\pm$ 0.03 & 1.0 $\cdot10^{-14}$\\
9--11 & 3388.16 $\pm$ 1.32 & 0.20 $\pm$ 0.01 & 1.5$\cdot10^{-14}$ \\
12--15 & 3388.15 $\pm$ 1.32 & 0.36 $\pm$ 0.03 & 1.7 $\cdot10^{-14}$\\
\hline
\end{tabular}
\end{minipage}
\end{table}



\section{Chromospheric and transition region properties of Proxima Centauri}

\subsection{Identification of chromospheric emission lines\label{sec_identification}}

\begin{figure*}
\begin{center}
\includegraphics[width=18cm]{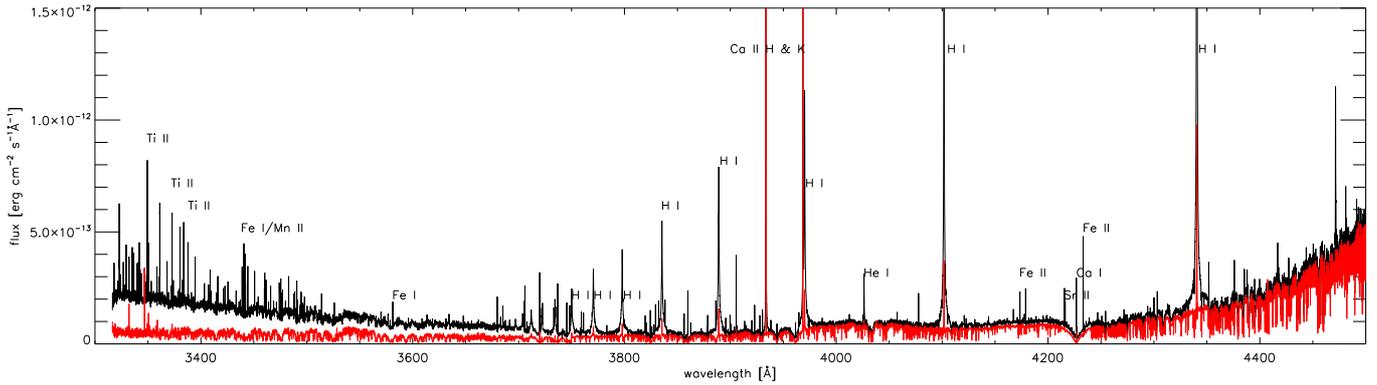}
\caption{\label{spectrum_blue} Main part of the UVES blue flare (black) and quiescent (red/grey)
spectrum. Several major chromospheric emission lines are identified in the plot. }
\end{center}
\end{figure*}

For the flare spectrum no. 9 in the blue arm and no. 76 in the red arm, respectively,
we produced an emission line list containing 474 chromospheric
emission lines, out of which 21 are from the red arm. In the blue arm, we analysed
the flare spectrum directly, while in the red arm we subtracted the quiescent
spectrum no. 74. The main part of spectrum no. 9 and spectrum no. 7  
can be found in Fig. \ref{spectrum_blue}.

To access
the line parameters, we fitted the 
background, central wavelength, half width $\sigma$, and line flux as free parameters
using a Gaussian profile for every manually indicated emission line
in a certain wavelength range. We typically chose 10 \AA\, as the wavelength 
range size for the fitting process, since the background variations for such a
short wavelength interval are negligible. The line fit parameters including a (possible) identification
can be found in Table \ref{linetable}. We only show a few rows of the Table
\ref{linetable} in this paper as an example, while the whole table
is provided only in electronic form.

\begin{table*}
\caption{\label{linetable} Ten rows of the line catalogue. The whole
table is accessible electronically.}  
\begin{minipage}{\linewidth}
\begin{tabular}[htbp]{cccclrcc}
\hline\hline
 central wavelength & half width & flux \footnote{ As discussed in section  \ref {UVES_observations}, the absolute errors for the given fluxes are about a factor of two.} & catalogued wavelength & ion & multiplet & id flag & comment\\
$[\AA]$ & [\AA] & [erg\,s$^{-1}$\,cm$^{-2}$] & [\AA]   & & & & \\
\hline \\[-3mm]
      3495.35 &   0.06  &    $8.70 \cdot 10^{-15}$  &   3495.285 &   \ion{Fe}{i}  &  238 &  2 &  blend 3495.370 CrII 2 \\
      3495.82 &   0.04  &    $1.25 \cdot 10^{-14}$  &   3495.831 &   \ion{Mn}{ii} &    3 &  0 & \\
      3496.80 &   0.03  &    $1.01 \cdot 10^{-14}$  &   3496.814 &   \ion{Mn}{ii} &    3 &  0 & \\
      3497.09 &   0.04  &    $3.93 \cdot 10^{-15}$  &   3497.115 &   \ion{Fe}{i}  &   78 &  2 &  em \\
      3497.51 &   0.03  &    $7.56 \cdot 10^{-15}$  &   3497.536 &   \ion{Mn}{ii} &    3 &  0 &  em \\
      3497.83 &   0.03  &    $1.21 \cdot 10^{-14}$  &   3497.843 &  \ion{ Fe}{i}  &    6 &  0 &  em \\
      3500.33 &   0.04  &    $8.34 \cdot 10^{-15}$  &   3500.340 &   \ion{Ti}{ii} &    6 &  0 & \\
      3500.87 &         &              &   3500.852 &   \ion{Ni}{i}  &    6 &  0 &  em, line not fitable \\
      3503.46 &   0.05  &    $7.53 \cdot 10^{-15}$  &   3503.474 &   \ion{Fe}{ii} &    4 &  0 & \\
      3504.89 &   0.02  &    $1.23 \cdot 10^{-15}$  &   3504.890 &   \ion{Ti}{ii} &   88 &  0 & \\
\hline
\end{tabular}
\end{minipage}
\end{table*}

The flux measurements given in the line catalogue may be affected by rather large errors
mainly for the following reasons. First, the description  with a Gaussian
may give a poor fit quality if the lines have broad wings (and should be
fitted with a Lorentzian/Voigt).
Second, the background/quasi-continuum may be  ill-defined. This is 
true for emission
lines in a broader emission line wing, but also for emission
cores of absorption lines and a variable background.
Moreover, nine Balmer/Paschen and other lines could not be fitted at all because they are too broad or
have an ill-defined background level (but the lines are clearly there). The absolute flux level
is also estimated to have rather large errors as discussed in section \ref{UVES_observations}. 

For the line identification, we  generally used the catalogue of \citep{Moore}.
A few lines were identified using the NIST database\footnote{Available online
under\\ \mbox{http://physics.nist.gov/PhysRefData/ASD/index.html}}. For the
identifications from the Moore catalogue, the multiplet is also given in Table \ref{linetable}. 
The spectra were first shifted to the
rest wavelength for the identification process which should enable us to
identify systematic line shifts caused by the flare. 

Three lines could not be identified with any known line.
Another 98 lines have only possible identifications for the following reasons:
(1) the line was not found in the Moore catalogue, (2) the wavelength shift to the 
possible laboratory wavelength is large, (3) the line is blended severely with 
other lines, and (4) (most often) the line is the
only one out of the multiplet.
Since we excluded doubtful features from our line list,
the list cannot be claimed to be complete, especially for weak lines.
The identified lines belong in total to 32 different ions. The number of lines found for the individual
ions is shown in Table \ref{identification}. 
Statistically, all lines are blue-shifted by 0.01 \AA. 

\setlength{\tabcolsep}{4.5pt}
\begin{table}
\caption{\label{identification}Chromospheric emission line identifications}  
  
\begin{tabular}[htbp]{ll|ll}
\hline\hline
ion & no. lines$ ^1$ & ion & no. lines$ ^1$  \\

\hline \\[-3mm]

 \ion{H}{i}  & 19(19)&  \ion{Mg}{i} & 10(6)\\
 \ion{He}{i} &  14(6) & \ion{K}{ii} & 1(0)  \\
\ion{Li}{ii} & 1(0) & \ion{Ca}{i} & 1(0)\\ 
\ion{Sc}{ii}& 7(4)& \ion{Ca}{ii}&  7(7) \\
 \ion{Ti}{i} & 3(0) & \ion{C}{i}  &  5(4) \\    
 \ion{Ti}{ii}&58(53)& \ion{O}{i}  &  3(3) \\  
 \ion{V}{ii} & 2(0)&  \ion{Ne}{i} &  1(0) \\    
 \ion{Cr}{i} & 9(7)&  \ion{Ne}{ii}  &  1(0) \\
 \ion{Cr}{ii}& 30(26)&   \ion{Al}{i} &  2(2)\\          
 \ion{Mn}{i} & 4(3)&   \ion{Si}{i} & 2(0)\\        
 \ion{Mn}{ii}&  9(8) &    \ion{Si}{ii}&  4(3) \\
 \ion{Fe}{i} &182(148)&  \ion{Cl}{ii}& 1(0)\\
 \ion{Fe}{ii}& 45(35)&   \ion{Ce}{ii}& 1(0)\\  
 \ion{Co}{i} & 5(2)&  \ion{Sr}{ii}&  2(2)\\                           
 \ion{Ni}{i} & 31(28)&   \ion{Zr}{ii}&  3(2)\\      
 \ion{Ni}{ii}&  5(5) &  \ion{Y}{ii} &  2(0) \\ 
\hline

\end{tabular}\\
$^{1}$ in brackets we indicate the number of secure identifications\\
\end{table}

We found Balmer lines up to number 19 during as well as outside the flare. The lines
are only stronger during the flare. This indicates that the densities during
the flare are not high enough to cause any Stark broadening of the lines, which would
lead to the merging of higher order Balmer lines \citep{Svestka}. This agrees well with
the electron densities measured using the X-ray data (see Subsection \ref{densitiesne}), which have
at most doubled during the flare compared to quiescence. 

We compare the line list to that of the large flare on CN~Leo \citep{CNLeoflare}, which consists of 1143 chromospheric
emission lines. The two catalogues overlap in the wavelength region from 3280 to 3850 \AA. In this range,
we compared the proportion of lines found in the CN~Leo flare to those found in this event. While for
\ion{Fe}{i}, \ion{Ti}{i}, \ion{Mn}{i}, \ion{Ni}{i}, and \ion{Cr}{i}, the proportion of lines found in the
CN~Leo flare compared to the lines found in the Proxima Centauri flare is (partially much) larger than 2, for the single
ionised species of the same elements the proportion is 1--1.5.   
 Moreover, nineteen lines in our present catalogue are not found in the CN~Leo flare. These are 4 \ion{He}{i}
lines, 2  \ion{Fe}{i}, 6 \ion{Fe}{ii}, 2  \ion{Zr}{ii} lines, 2  \ion{Co}{i}, 1  \ion{V}{ii}, 1  \ion{Ti}{i},
and 1  \ion{H}{i}. This seems to indicate that the flare energy is deposited very effectively
in the higher chromosphere/lower transition region, where helium and Balmer lines are formed, and
also the lines of singly ionised metals should be formed.   


\subsection{Line asymmetries}\label{asymmetries}

During the flare, line asymmetries are found in both the blue and the red spectra.
The Balmer lines, \ion{He}{i}, and \ion{Ca}{ii} K show additional flux in the
red wing. In the blue arm, we investigated the Balmer lines H$\gamma$, H$\delta$,
and the Balmer lines at 3889~\AA\, and 3835~\AA. We refrained from fitting H$\epsilon$
because of the heavy blending with the \ion{Ca}{ii} H line. We also failed to fit the higher
Balmer lines because of the extreme width of these lines during the flare, the blending
with multiple metal lines, and the low line amplitudes of the very highest Balmer lines.
Unfortunately, all \ion{He}{i} lines in the blue arm are weak or blended and
decay very fast, so that in the spectrum no. 10 after the flare peak spectrum no. 9,
the helium lines have almost vanished. The \ion{Ca}{ii} K line being an unblended strong metal
line shows a clear red wing asymmetry. No other metal lines have
clear asymmetries. In the red arm, we find an asymmetry for the \ion{He}{i}
line at 6678 \AA\, in the flare peak spectrum no. 76, while in spectrum no. 77,
the line has already faded  too much to display  any asymmetry. 
The other three helium lines in the red arm spectrum are either
too weak or too complicated in their structure (blend with another helium line)
to show unambiguous asymmetries. Hence, the only line for which we could study the
asymmetries  with higher time resolution of the red arm spectra
is H$\alpha$. This line shows indeed that the asymmetries have a strong
time dependence that cannot be studied with the time resolution of the blue
arm spectra. We fitted the line asymmetries with two Gaussians: one for the
main (narrow) line component and one for the (broad) wing component. For all
fits, we subtracted the quiescent spectrum directly before the flare. 
The shifts between the broad and the narrow Gaussian are listed in Table
\ref{shifts} for various lines and for the H$\alpha$ line with higher time
resolution in Table \ref{Hashifts}. Although the formal errors found in the
fitting process for the line shifts are small, we assume a ten percent error
for all the line shifts. 

\begin{figure*}
\begin{center}
\includegraphics[width=16cm]{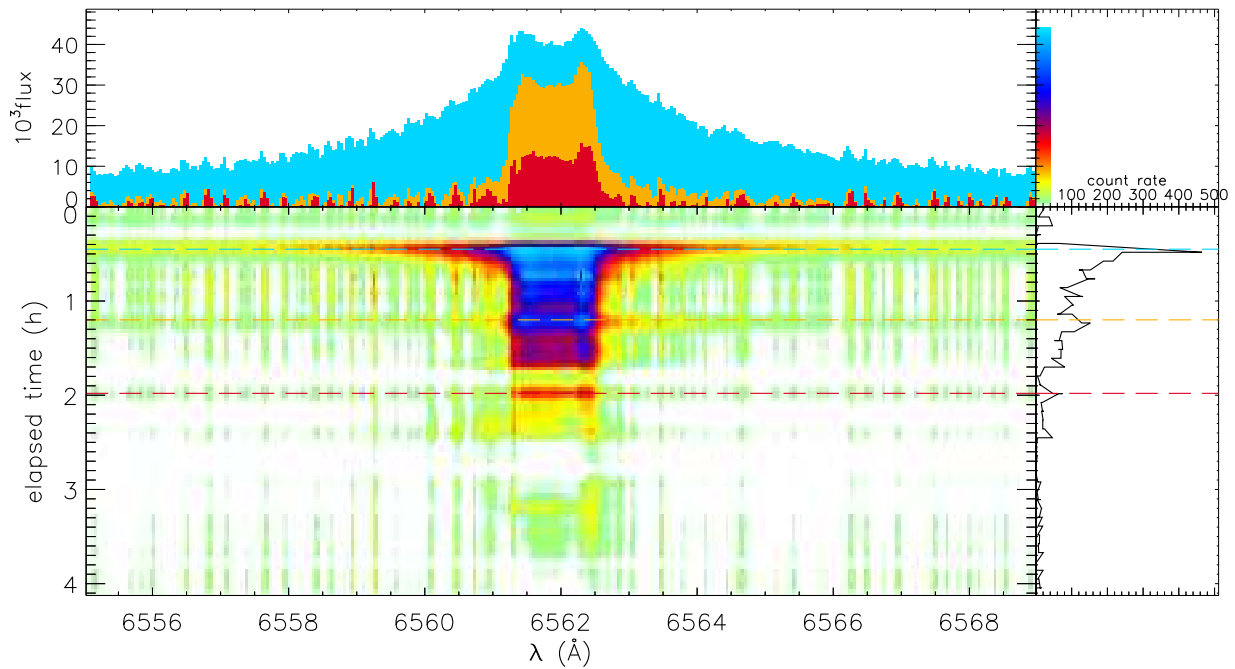}
\caption{\label{Halpha_map} The H$\alpha$ line variations in time with the quiescence spectrum no. 74
subtracted. The top panel 
shows three individual spectra as marked by the horizontal dashed lines in the
time map below. On the right hand panel, we show the light curve of the 
integrated flux of the quiescence subtracted spectrum for the shown 
wavelength region. The time is measured in elapsed time
since 5:50 UT. }
\end{center}
\end{figure*}

The high time resolution of H$\alpha$ clearly shows that the asymmetries are a  dynamical phenomenon as can be 
seen in Fig. \ref{Halpha_map}. The asymmetries for the first secondary flare can
be identified in the time map itself (elapsed time is about 1.3 h). In addition, it can be seen in the
top panel of Fig. \ref{Halpha_map}, where the yellow shaded spectrum shows the most prominent asymmetry in the wing.
 The third peak in this light curve also shows broad wings in the
H$\alpha$ spectra,  which are, however, weaker, thus could not be fitted.  Starting with 
spectrum no. 149 (08:52:10 UT) and no. 150 with a blue asymmetry, the asymmetry turns to
the red side in spectrum no. 152 and 153 (09:00:48 UT).  These asymmetries can also be noted
in the time map (elapsed time about 3.2 h in Fig. \ref{Halpha_map}).
 While in the
blue spectra there are only red asymmetries, in H$\alpha$ there are also blue ones.
Indeed for the entire integration time of the blue spectrum no. 9, there are mostly
blue-shifts for H$\alpha$; only for red spectrum no. 76 corresponding to the first few minutes  
exposure time of spectrum no. 9 there is 
a red shift (the spectrum no. 9 corresponds to elapsed time from 0.4 to 0.7 h in Fig. \ref{Halpha_map}).
Since the amplitude decays exponentially, the conditions at the
beginning of the spectrum should dominate the spectrum. Therefore, blue spectrum
no. 9 and spectrum no. 76 yield similar red shifts. After red spectrum no. 80 (corresponding to an elapsed time of about 0.6 h),
the asymmetries and large wings disappear -- but re-appear in red spectrum
no. 97 -- 102 (corresponding to an elapsed time of about 1.1 to 1.3 h in Fig. \ref{Halpha_map}). This agrees with the  red wing asymmetries in blue spectrum no. 10, although
the asymmetry in the red arm is invisible for the whole integration time of this spectrum.
That the time scales on which the asymmetries evolve can be very short can be seen
in red spectrum no. 75 and blue spectrum no. 8. While the red spectrum  shows
a clear blue-shift, in the blue spectrum the lines either cannot be  fitted
or show no real line shifts, but do display broad wings (see Tables \ref{shifts}, \ref{Hashifts}).

We interpret all these line shifts as manifestations of mass motions in the chromosphere
as e.\,g. \citet{CNLeoflare}. The high velocity blue-shift at the flare onset corresponds
to evaporation of chromospheric material, while during the decay phase raining down as well as
rising material can be found.  

\begin{table}
\caption{\label{shifts}Line shifts of the broad line component for 
diverse lines from the flare-only spectra. }  
\begin{tabular}[htbp]{cccc}
\hline
\hline
line & velocity & velocity & velocity\\
 & [km\,s$^{-1}$] & [km\,s$^{-1}$] & [km\,s$^{-1}$]\\
     & spec. no. 8(75) & spec. no. 9(76) & spec. no. 10(77) \\
     & 05:47:14 & 06:14:39 & 06:45:12 \\
\hline \\[-3mm]
H$\gamma$ (\@ 4340 \AA) & -- & 4.8  & 12.4  \\
H$\delta$ (\@ 4101 \AA) & 3.9  & 4.9 & 10.2 \\
\ion{H}{i} \@ 3889 \AA & -- & fit ambiguous & 13.1  \\
\ion{H}{i} \@ 3835 \AA & 2.1  & fit ambiguous & 9.4 \\
\ion{Ca}{ii} K & -- & 5.1 & 21.4 \\
\ion{He}{i} \@ 6678 \AA & -- & 6.0 & -- \\
\hline
\end{tabular}
\end{table}

\begin{table}
\caption{\label{Hashifts}Line shifts of the broad line component for 
the H$\alpha$ line from the flare-only spectra. }  
\begin{tabular}[htbp]{cccccc}
\hline
\hline
spectrum  & start time  & velocity & spectrum  & start time  & velocity \\
no. & (UT) & [km\,s$^{-1}$] & no. & (UT) & [km\,s$^{-1}$] \\
\hline \\[-3mm]
75 & 06:12:16 & -44.3  & 97 & 06:59:50 & 8.7 \\
76 & 06:14:29 & 5.3   &  98 &  07:02:00 & 13.7 \\
77 & 06:16:37 & -10.5  & 99 &  07:04:09 & 13.7 \\
78 & 06:18:48 & -9.6 & 100 & 07:06:19 & 33.7 \\
79 & 06:20:57 & -5.9  & 101 & 07:08:28 & 32.9 \\
80 & 06:23:06 & 1.4   & 102 & 07:10:37 & 3.7 \\
\hline
\end{tabular}
\end{table}

\subsection{Theoretical modelling of the flare with PHOENIX}

We modelled the flare spectra with theoretical PHOENIX chromospheric
flare spectra. These spectra were originally computed to fit the mega-flare on CN~Leo
\citep{CNLeomodel}. The stars CN~Leo and Proxima Centauri have  similar photospheric
properties. While for Proxima Centauri, $T_{\mathrm{eff}}$ is about 3100 K, for CN~Leo a
$T_{\mathrm{eff}}$ of 2900 K was used for the underlying photosphere in the model calculation.
\citet{Short1} studied the influence of the underlying photosphere on a
quiescent chromospheric model and found that a difference of 200 K in $T_{\mathrm{eff}}$
leads to uncertainties in the parameters of the column mass at the onset of the transition 
region as well as the column mass of the temperature minimum of about 0.3 dex.
The finest step in our flaring model grid is 0.2 dex. Since we use a photosphere
with a lower $T_{\mathrm{eff}}$ than is realistic, we should infer a too high  column mass for the onset of
the transition region and the temperature minimum.   
Details about the model construction can be found in \citet{CNLeomodel}. 
In this paper, we constructed the flaring model as a linear combination of a quiescent spectrum observed
directly before the flare and a  flaring model spectrum.

\begin{figure}
\begin{center}
\includegraphics[width=8cm]{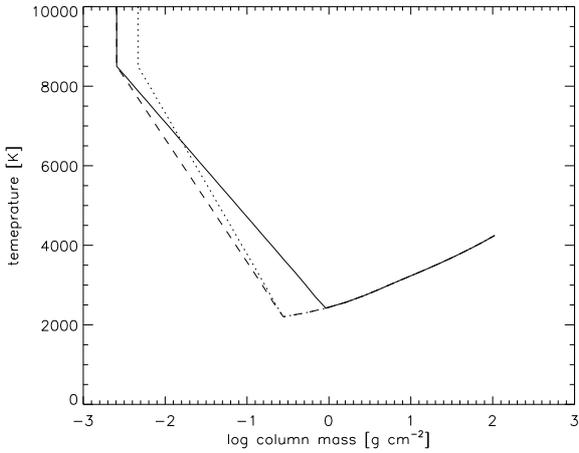}
\caption{\label{flaremodel_graph} Example of the temperature structure of the best-fit flare models for
different spectra: black line: red spectrum no. 76 (flare onset); dotted line: red spectrum no. 97 
(secondary flare); dashed line: blue spectrum no. 9 }
\end{center}
\end{figure}

\begin{figure}
\begin{center}
\includegraphics[width=8cm]{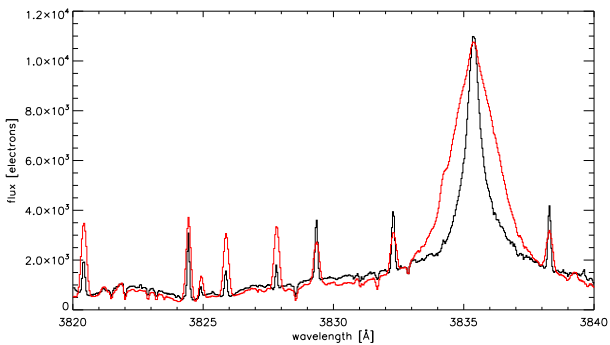}
\caption{\label{phoenix} Exemplary comparison between observed spectrum (black)
and our best-fit model spectrum (red) for blue spectrum no. 9. }
\end{center}
\end{figure}

For the red and the blue arm respectively, we then used a global normalisation of the spectra and a
grid of different filling factors of the flare model spectra to the quiescent observed
spectra to help us find the best fit to the observed flare spectra. The filling
factors used for the quiescent spectra are 0.85, 0.90, 0.93, 0.95, 0.96, 0.97, 0.98, 0.99, 0.995, and  0.999.
The best fit was found using a $\chi^2$ analysis again following \citet{CNLeomodel}.
The mean best-fit parameters for the five best-fit flare models can be found in
Table \ref{flaremodel} with the standard deviation of the mean. Examples of the temperature
structure of the best-fit models for a selection of different spectra can be found in Fig. \ref{flaremodel_graph}.
For a small wavelength range in the blue, a comparison between the observed spectrum and
best-fit model spectrum can be found in Fig. \ref{phoenix}.

\begin{table}
\caption{\label{flaremodel}Mean best-fit flare model parameters in time. }\scriptsize  
\begin{tabular}[htbp]{cccccc}
\hline
\hline
spec.  & T$_{chrom}$  & log & log & grad TR & filling \\
no.    & [K]  &  $cmass_{Tchrom}$ & $cmass_{Tmin}$ & &factor\\
\hline \\[-3mm]
76 & 8400 $\pm$ 300 & -2.5 $\pm$ 0.2 & -0.3 $\pm$ 0.2 & 11.0 $\pm$ 0.1 & 2.0 $\pm$ 0.5\\
77 & 8400 $\pm$ 200 & -2.5 $\pm$ 0.1 & -0.3 $\pm$ 0.2 & 11.1 $\pm$ 0.3 & 2.0 $\pm$ 0.1\\
78 & 8300 $\pm$ 200 & -2.7 $\pm$ 0.3 & -0.4 $\pm$ 0.0 & 11.1 $\pm$ 0.3 & 2.0 $\pm$ 0.1\\
80 & 8200 $\pm$ 300 & -2.7 $\pm$ 0.3 & -0.4 $\pm$ 0.2 & 11.1 $\pm$ 0.3 & 2.0 $\pm$ 0.1\\
82 & 8300 $\pm$ 300 & -2.9 $\pm$ 0.2 & -0.9 $\pm$ 0.3 & 11.0 $\pm$ 0.4 & 4.0 $\pm$ 0.1\\
\hline \\[-3mm]
95 & 8000 $\pm$ 100 & -3.0 $\pm$ 0.1 & -1.1 $\pm$ 0.3 & 11.0 $\pm$ 0.4 & 6.0 $\pm$ 1.0\\
96 & 8400 $\pm$ 300 & -2.6 $\pm$ 0.3 & -0.2 $\pm$ 0.2 & 11.0 $\pm$ 0.1 & 2.0 $\pm$ 0.8\\
97 & 8400 $\pm$ 300 & -2.5 $\pm$ 0.2 & -0.3 $\pm$ 0.2 & 11.0 $\pm$ 0.1 & 1.5 $\pm$ 0.4\\
98 & 8400 $\pm$ 300 & -2.7 $\pm$ 0.2 & -0.3 $\pm$ 0.2 & 11.1 $\pm$ 0.3 & 1.5 $\pm$ 0.4\\
99 & 8400 $\pm$ 300 & -2.7 $\pm$ 0.2 & -0.3 $\pm$ 0.2 & 11.1 $\pm$ 0.3 & 1.5 $\pm$ 0.4\\
101& 8200 $\pm$ 300 & -3.0 $\pm$ 0.1 & -1.1 $\pm$ 0.5 & 10.9 $\pm$ 0.3 & 4.0 $\pm$ 0.1 \\ 
\hline \\[-3mm]
9  & 8500 $\pm$ 100 & -2.5 $\pm$ 0.2 & -0.3 $\pm$ 0.2 & 11.0 $\pm$ 0.4 & 0.8 $\pm$ 0.1\\
10 & 8400 $\pm$ 300 & -2.9 $\pm$ 0.2 & -0.8 $\pm$ 0.3 & 11.0 $\pm$ 0.1 & 0.9 $\pm$ 0.1\\
\hline
\end{tabular}
\end{table}

The model parameters of the fitting process from the blue and the red
arm agree  well with each other with both exhibiting a shift towards
flare models further out in the atmosphere later in the decay phase.
For red models later than no. 82, no meaningful fit could be found with
many models exhibiting nearly the same $\chi^2$. Even later during the 
secondary flare phase, a fitting of the flaring models also turned out to
be possible again, starting at red spectrum no. 95. The secondary flare can be
seen in the models as a re-shift to higher-pressure chromospheric flare models.
For the higher-pressure models, a higher temperature for the onset of
the transition region is also found.

The only disagreement between model parameters can be found for the filling factor. Comparing
red spectrum no. 76 to blue spectrum no. 9 shows that the blue spectrum 
has a  lower filling factor. This may indicate that there is an error in the
normalisation, which was done for the red arm where the
influence of the flare on the continuum should be small and then applied to
the blue arm. Another possibility of the different filling
factors is a height dependence of the filling factor.
The blue arm with its wealth of metal lines should trace the mid-chromosphere,
while the red arm with the \ion{He}{i}, Balmer, and Paschen lines should
trace the upper chromosphere. The filling factor found for the red
spectra models of about two percent agrees  well with that
found by the EM-T diagnostics from the X-ray data, which is  three percent (see section \ref{loopEMT}).


\section{Discussion}

The multi-wavelength character of the observations allows us to characterise
general behaviour and energetics of the flare for various wavelength bands, hence
 different temperature regimes and also cross-check these using different
instruments. \\

\subsection{Flare energetics and excess emission}

The flare exhibited a broadband excess emission in the 
far blue of the UVES spectral arm, which can be easily identified in Fig. \ref{spectrum_blue}.
A similar excess was found for larger flares and could in these cases
be ascribed to blackbody emission \citep{CNLeoflare, Hawley_Fisher}. In the 
flare observed here, it was impossible to fit the excess emission with
a blackbody model. We nevertheless calculated the excess flux, noting that the
shape of the spectrum suggests that the bulk of the broadband emission is
not covered by the UVES spectrum. We integrated the flare spectrum up to 3600~\AA\, and subtracted (1) the quiescent flux, (2) the general
flux increase seen in the blue arm during the flare,
and (3) the chromospheric line flux in this wavelength interval for the flux excess calculation. 
 We also calculated the total of the 
chromospheric line fluxes in the blue arm and in H$\alpha$. We compared the H$\alpha$ line flux  to the value
one obtains using the $\chi$ method introduced by \citet{chi}, which is based only on the equivalent width of the
line, the bolometric luminosity, and the distance of the star (so one needs no absolute flux calibration). Using the
$\chi$ value of $0.176\cdot 10^{-4}$  for an M6 dwarf, $L_{bol}$ of $6.7\cdot 10^{30} \mathrm{erg\,s^{-1}}$, and a measured peak EW of 3.4, we evaluated the same flux of 
2.0 $\cdot$10$^{-12}$ erg\,s$^{-1}$\,cm$^{-2}$ as in our flux 
calibration  (or 1.8 $\cdot$10$^{-12}$ erg\,s$^{-1}$\,cm$^{-2}$ using $L_{bol}$ of $6.0\cdot 10^{30} \mathrm{erg\,s^{-1}}$). 
The various fluxes can be found in Table \ref{fluxes} together with the 
resulting energies assuming a filling factor for the flare of two percent for the time range of the blue UVES spectrum no. 9 
from 6:14 to 6:45 UT. \\

\begin{table}
\caption{\label{fluxes}Various fluxes and flare energetics for different wavelength regimes. }  
\begin{minipage}{8cm}
\begin{tabular}[htbp]{cccc}
\hline
\hline
\multirow{2}{*}{Wavelength band} & \multirow{2}{*}{Instrument} & Flux \footnote{The errors in the UVES fluxes are about a factor of two (see section \ref{UVES_observations}).}& Energy \\
& & [erg\,s$^{-1}$\,cm$^{-2}$] & [erg] \\
\hline \\[-3mm]
chrom. em. lines & UVES blue arm & 1.2$\cdot$10$^{-11}$ & 8.5$\cdot$10$^{28}$\\
H$\alpha$ emission  & UVES red arm & 2.0$\cdot$10$^{-12}$ & 1.6$\cdot$10$^{28}$\\ 
blue excess em. & \multirow{2}{*}{UVES blue arm} & \multirow{2}{*}{5.3$\cdot$10$^{-10}$} & \multirow{2}{*}{3.8$\cdot$10$^{30}$}\\
(3300-3600 \AA) & & &\\
X-rays & \multirow{2}{*}{PN+RGS \emph{XMM}} & 2.2$\cdot$10$^{-11}$ & \multirow{2}{*}{1.9$\cdot$10$^{29}$}\\
(0.2-10 keV) & &$\pm 0.2 \cdot$10$^{-11}$ & \\
\hline
\end{tabular}
\end{minipage}
\end{table}

\subsection{H$\alpha$ self-absorption}\label{self-absorption}

A comparison to the  flux-calibrated
spectrum of Proxima Centauri presented by \citet{Cincunegui} shows that the H$\alpha$
line amplitude in our spectra is extremely small during the whole three days of
our observations. The line amplitude during the flare is comparable to the
line amplitude of the spectrum from \citet{Cincunegui}, which was taken during a quiescent 
state. This 
indicates that Proxima Centauri's chromosphere was generally  in a low activity state 
during the observations. The occurring flare enhanced the H$\alpha$ line flux only to a
level that can also be  found during a quiescent state (at other times). This fits well with the 
relatively low X-ray flux found in our observations (see section \ref{quasi-quiescent}).
 
The H$\alpha$ line is heavily self-absorbed during our observations. This self-absorption
is in most cases asymmetric, but the line peak at the red side is  higher
than at the blue; this peak asymmetry should not be confused with the wing asymmetry
described  in section \ref{asymmetries} of this paper. Two examples can easily be noted in the 'flare only' spectra shown
in the top panel of Fig. \ref{Halpha_map}: while for the flare spectrum the two peaks are 
about of the same height and the line centre therefore symmetric, for the two spectra from the decay phase
the red peak is higher than the blue peak.  We note that the spectra of \citet{Cincunegui} have lower spectral resolution 
hiding a possible self-reversal. 
We therefore compared our data to processed HARPS data from the ESO archive
facility\footnote{Based on data obtained from the ESO Science Archive Facility
from program 072.C-0488(E).}, which cover the spectrum of Proxima Centauri from
3782 to 6912~\AA, in the time from 2004 to 2007. These spectra have not yet been
published  and show a wide variation in the H$\alpha$ line amplitude,
with the spectra from \citet{Cincunegui}  at the high end of measured H$\alpha$ line amplitudes and our observations
at the low end. Like our spectra the HARPS spectra exhibit in most cases a self-reversal with more 
flux in the red side peak.

We checked the red peak asymmetries using our models, which also display
H$\alpha$ self-absorption at the line centre that is in all cases symmetric. This
would be expected for one-dimensional (1D) chromospheric models without mass motion, where the self-reversal
is a pure non-local thermal equilibrium (NLTE) effect. \citet{Allred}
computed 1D hydro-dynamical models of M dwarf flares. For various evolutionary stages of the flare in
 their simulations, the H$\alpha$ line profiles
show 
the red peak asymmetries also observed in our and the HARPS spectra. In
the hydro-dynamical models, the
peak asymmetry is caused  by mass motions during the simulated flare. It appears reasonable
to assume that similar mass motions are also common in the quiescent state to explain the
observed peak asymmetries. \\

\subsection{Magnetic field}

The dip in the magnetic field coincides in time with the flare in
the third night. We estimated the energetics as a consistency check, in case a physical connection
were possible.
The level of the magnetic field changes from $\approx 300\pm100$~G just before the flare 
to $\approx 100\pm100$~G at the time the flare starts in the optical. So, the nominal change in mean magnetic field is $200$~G, although the errors are very large. As a consistency check however, we can calculate that this corresponds to a released magnetic energy density of 

$\eta = \frac{B^2}{8\pi} = 1.6\times 10^3$~erg~cm$^{-3}$.

In section~\ref{loopEMT}, we calculated the flaring volume to be 
$V_{flare} \approx 6.25 \times 10^{29}$~cm$^3$, yielding a total released 
magnetic energy of $1.0\times10^{33}$~erg. We computed the total X-ray energy released 
during the first 6.5~ks of the flare to be $1.5\times 10^{31}$~erg by integrating the flare and quiescent spectral models from Table~\ref{3temp} over the energy range 0.2-10~keV and calculating the difference between the two. This is two orders of magnitude less than 
our estimate for the total released magnetic energy. However, since the detailed loop configuration 
is not known and we simply estimated the flaring volume as $V = L^{3}$ with $L$ being 
the loop half length, the volume itself can only be an order-of-magnitude estimate. 
Therefore, we interpret the derived numbers as a general confirmation that the emitted 
X-ray energy during the flare is within the energy budget of the theoretically released magnetic energy.
The question, of whether the dip in magnetic field is really physically connected to the flare 
must unfortunately remain unsolved.

Other flare stars of similar spectral type do show higher quiescent activity levels. Although
magnetic field measurements of mid and late-type M dwarfs are still
quite rare, there is a time series measurement for the M5.5 star
CN~Leo \citep{CNLeoBf} with mean magnetic field of 2.2 kG covering a large flare. A 
dip in $Bf$ at the time of the flare is also seen in these data, although there are other field variations of 
similar amplitude that are not 
associated with flares. Thus, it remains unclear whether the change in magnetic field is associated with the flare.

\citet{CNLeoBf} found a correlated change in the distributions of magnetic field $Bf$ and H$\alpha$ emission in 
their data of CN~Leo. We searched for a similar correlation in our data, but found no one, which 
is unsurprising as the magnetic field changes are only small. We note that the H$\alpha$ emission
during the course of these observations are quite low (see section \ref{self-absorption}) corresponding to a
weak magnetic field. Unfortunately, the observation of \citet{ReinersBf}, which found a stronger magnetic field
for Proxima Centauri (as described in section \ref{avmagfield}) does not include the H$\alpha$ line. \\

\subsection{Line asymmetries}

Line asymmetries during flares have often been found in high
resolution spectra mostly for \ion{H}{i} and \ion{He}{i} lines. Examples are
red line asymmetries for
a large flare on LHS~2034 (M6.5)  \citep{LHS2034},  blue line 
asymmetries for a flare on AT~Mic (M4) \citep{Gunn}, and red line asymmetries for  
flares on AD~Leo (M3.5Ve) \citep{Crespo} (for further examples see also references therein). 
For the CN~Leo
mega-flare described by \citet{CNLeoflare}, blue-wing as well as red-wing
asymmetries have also been found. Asymmetries were normally ascribed
by these authors to mass motions, which were partially modelled with multiple kernels and compared
to the chromospheric downward condensations (CDCs) observed in the Sun in the case
of the red asymmetries. During the CN Leo flare,  asymmetries in the wings
of \ion{Ca}{ii} lines were also found for the first time. Surprisingly, we found
asymmetries in the  \ion{Ca}{ii} H and K line for this much smaller
flare on Proxima Centauri, though \citet{Crespo} found that larger flares
have stronger asymmetries and found no asymmetry in the \ion{Ca}{ii} H and K line
in their flare data (their longest flares lasted about 30 minutes).

In contrast to the CN~Leo mega-flare we also found blue wing asymmetries
during flare decay and for the first time a re-appearance of
the asymmetries during the secondary flare phase. 
The vanishing and re-appearance of the broad wings combined with the asymmetries
of the H$\alpha$ line coincides with the secondary flare process seen in the X-ray light curve
(see Fig. \ref{lightcurve3} and Table \ref{Hashifts}). When the secondary flare started, the broad component vanished,
but reappeared some time later. The reappearance of the broad wings can also
be identified in the H$\alpha$ light curve (see Fig. \ref{lightcurve3}) as the peak after the
main flare peak. 
The data again illustrates
the  dynamic behaviour of the asymmetry phenomenon, which provides additional evidence of
 mass motions.\\

\subsection{Theoretical modelling}

The clear secondary flare found by the flare models for the chromosphere strengthens
our confidence in this approach of mixing quiescent observations with flaring
model spectra. However, the general caveats of the method presented  by
\citet{CNLeomodel} still apply: the 1D models show (for simplicity reasons) a linear temperature rise,
which can explain part of the difficulties in fitting large wavelength regions
with the same model spectrum equally well. Moreover, the 1D
models are a very rough description of the horizontally as well as vertically
highly complex temperature pattern (including shocks) of the (solar) chromosphere. Therefore,
more physical chromospheric models should include hydrodynamic simulations 
and dynamic ionisation in either 1D or 3D.\\

\begin{figure}
\begin{center}
\includegraphics[width=8cm]{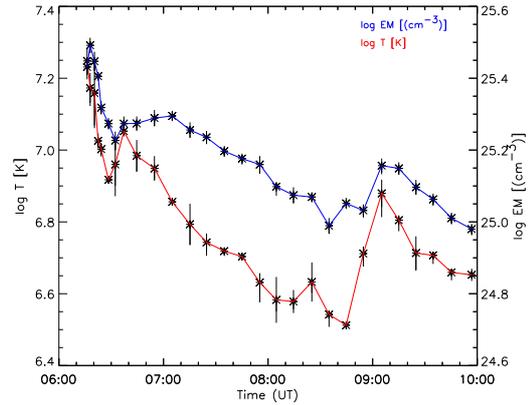}
\caption{\label{em_evol_temp} Temporal evolution of flare temperature and emission measure.}
\end{center}
\end{figure}

\begin{figure}
\begin{center}
\includegraphics[width=8cm]{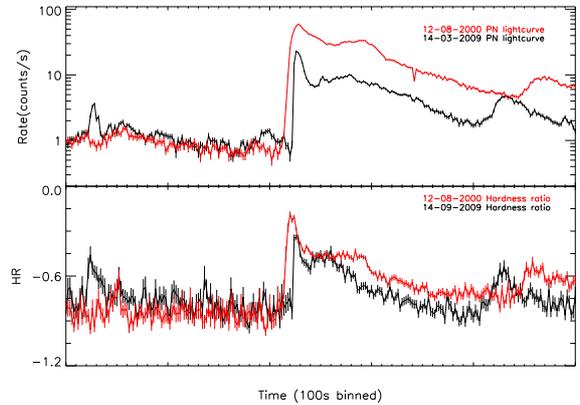}
\caption{\label{comparison}  Light curves with 100s binning and hardness ratio of flares observed 
on Proxima Centauri by \emph{XMM-Newton} on 12 August 2001 (red) and 14 March 2009 (black), respectively. }
\end{center}
\end{figure}
   
\subsection{Secondary flare events}

The nature of this flare with its two secondary events following the first peak in the light curve 
raises the question of what kind of magnetic structure we are dealing with. There are two 
possibilities: it could be a single loop that experiences two additional heat pulses after the 
initial flare, or an arcade structure where several loops light up in sequence. 
The heat pulse refers to the process where the electron beam is accelerated 
in the coronal part of the magnetic flare loop, then propagates along the magnetic 
field lines down to the chromosphere,
where the electrons are fully thermalised. A heat pulse represents the heating of the atmosphere by the electron beam.
\citet{Reale_ProxCen_2004} interpreted an earlier long-duration flare light curve of 
Proxima Centauri as being produced by two loops flaring in sequence, since the coronal temperature 
during the second peak was flat-topped and could not be adequately explained by a second heat pulse in the same 
loop. In our data however, the temperature and emission measure evolution follows the trend seen 
in the light curves closely as can be seen comparing Fig. \ref{em_evol_temp} and Fig. \ref{lightcurve3}. 
 
From our analysis of the EM-T diagram for this flare, we derive very similar EM-T slopes for all 
three flare decays, each one indicating that a moderate amount of continuous reheating is present. 
The thus derived loop lengths are of the same order of magnitude. Even if the errors are rather 
large, there is a decreasing trend discernible within these three lengths; the lengths derived 
from the second and third peaks are comparable, while the length derived from the first peak
 seems to be larger by a factor of roughly two. This, and that the amount of 
reheating during the decays is very similar, suggests that we may see a series of similar loops 
in an arcade lighting up in sequence. However, owing to the large errors in the loop lengths, the 
possibility of a single loop with several heat pulses cannot be ruled out completely. A comparison
of the X-ray and hardness ratio light curves for the mega-flare observed by \citet{Reale_ProxCen_2004} 
and \citet{Guedel_ProxCen_2}
and the flare presented here can be found in Fig. \ref{comparison}. The similarity of the light curves
of the two events (despite the different count rates) is quite remarkable.


\section{Summary and conclusions}

We have presented multi-wavelength observations of the different activity stages of Proxima
Centauri with particular emphasis on a long duration flare. We have tried to construct an overall picture of the
atmosphere from the photosphere to the corona. We have determined a low average magnetic  field,
corresponding to low average X-ray and H$\alpha$ emission, with the latter being heavily self-absorbed. For the flare, we
have compiled a chromospheric emission line list and found asymmetries in the broad wings of Balmer, \ion{He}{i},
and \ion{Ca}{ii} H \& K lines, which we ascribe to mass motions, which could explain coronal abundance
changes. From the X-ray data, we have determined
coronal densities, abundances, emission measures, and temperatures. The higher emission measure
of the low temperature component  and the higher Fe abundance during the flare
fit well with the detection of a forbidden optical \ion{Fe}{xiii}
line during this time. 
The flare  
light curve is similar to the mega-flare for the same star described by
\citet{Reale_ProxCen_2004} and \citet{Guedel_ProxCen_2}  exhibiting two bumps during
the decay phase. The secondary flares are of a  similar loop length as the
main flare, show line asymmetries, and chromospheric flare modelling
shows that the first one exhibits 
similar chromospheric parameters as the main one. All this indicates that the
events are not  independent of each other, but result from the same loop or
at least an arcade with several arcade loops igniting consecutively.   
This seems to be a  typical flare scheme for Proxima Centauri given, that
it hs been observed for the second time, and  \citet{Reale_ProxCen_2004}
noted the similarity to solar flares. In contrast to Proxima Centauri, CN Leo,  another
well-observed M5.5 but more active dwarf never showed a comparable flare cascade
during a long-duration flare. On the other hand, another highly active dM4.5e star, YZ~CMi,
also exhibited  a series of secondary flares during a white light mega-flare as described
by \citet{Kowalski_1,Kowalski_2}, who also speculated that the flare on YZ CMi originated in
a complex arcade with a sequence of reconnecting loops. If these interpretations are
correct, mid-type M dwarfs seems to have a flaring loop geometry that may be similar to the 
Sun, which will help to place constraints on turbulent dynamo theories. 

\begin{acknowledgements}
S.~L. acknowledges funding by the DFG in the framework
of RTG 1351. K.~P. acknowledges funding
under project number DLR 50OR0703. 
N.~R. acknowledges financial support by the DLR under project no. 50OR1002.  
\end{acknowledgements}

\bibliographystyle{aa}
\bibliography{papers}

\end{document}